# Highway Managed Lane Usage and Tolling for Mixed Traffic Flows with Connected Automated Vehicles (CAVs) and High-Occupancy Vehicles (HOVs)

**Max T.M. Ng**
Graduate Research Assistant
Transportation Center
Northwestern University
600 Foster Street
Evanston, IL 60208, USA
Email: maxng@u.northwestern.edu

**Hani S. Mahmassani***
William A. Patterson Distinguished Chair in Transportation
Director, Transportation Center
Northwestern University
600 Foster Street
Evanston, IL 60208, USA
Email: masmah@northwestern.edu
Tel: 847.491.2276

* Corresponding Author

## ABSTRACT

This paper investigates managed lane (ML) toll setting and its effect under mixed traffic of connected automated vehicles (CAVs), high-occupancy vehicles (HOVs), and human-driven vehicles (HDVs), with a goal to avoid flow breakdown and minimize total social cost. A mesoscopic finite-difference traffic simulation model considers the flow-density relationship at different CAV market penetration rates, lane-changing behavior, and multiple entries/exits, interacting with a reactive toll setting mechanism. The results of the Monte Carlo simulation suggest an optimal policy of untolled HOV/CAV use with HDV tolls in particular scenarios of limited CAV market penetration. Small and targeted tolling avoids flow breakdown in ML while prioritizing HOVs and other vehicles with high values of time. Extensions of the formulation and sensitivity analysis quantify the benefits of converting high-occupancy HDVs to CAVs. The optimal tolling regime combines traffic science notions of flow stability and the economics of resource allocation.

# 1 INTRODUCTION

High-occupancy lanes, initially designated for high-occupancy vehicles, became available under certain conditions for use by low-occupancy vehicles willing to pay a toll since about 1995 (*1*). Such lanes are also known as high-occupancy toll (HOT) lanes or, more generally, managed lanes (MLs), while the other general purpose lanes (GPLs) remain available for all traffic. The tolling mechanism applied in these HOT lanes/MLs may be predicated on various objectives, including maximum flow and minimum social cost, and has remained a topic of interest to economists, traffic planners, and both public and private infrastructure operators.

In the meantime, connected and autonomous vehicles (CAVs) enable centralized and decentralized coordination among vehicles running on highways, and their role in optimizing traffic flow was also explored. Talebpour et al. (*2*) showed the potential beneficial effect of a reserved lane for CAVs, which was however minimal at lower market penetration rates (MPRs) relative to the available capacity (number of lanes). Therefore, assigning a dedicated CAV lane may induce additional traffic congestion in GPLs. This motivated further research, e.g. (*3*, *4*), to combine a dedicated CAV lane with a human-driven vehicle (HDV) toll lane with models of routing and planning for mixed autonomous flows in toll lanes in a network setting.

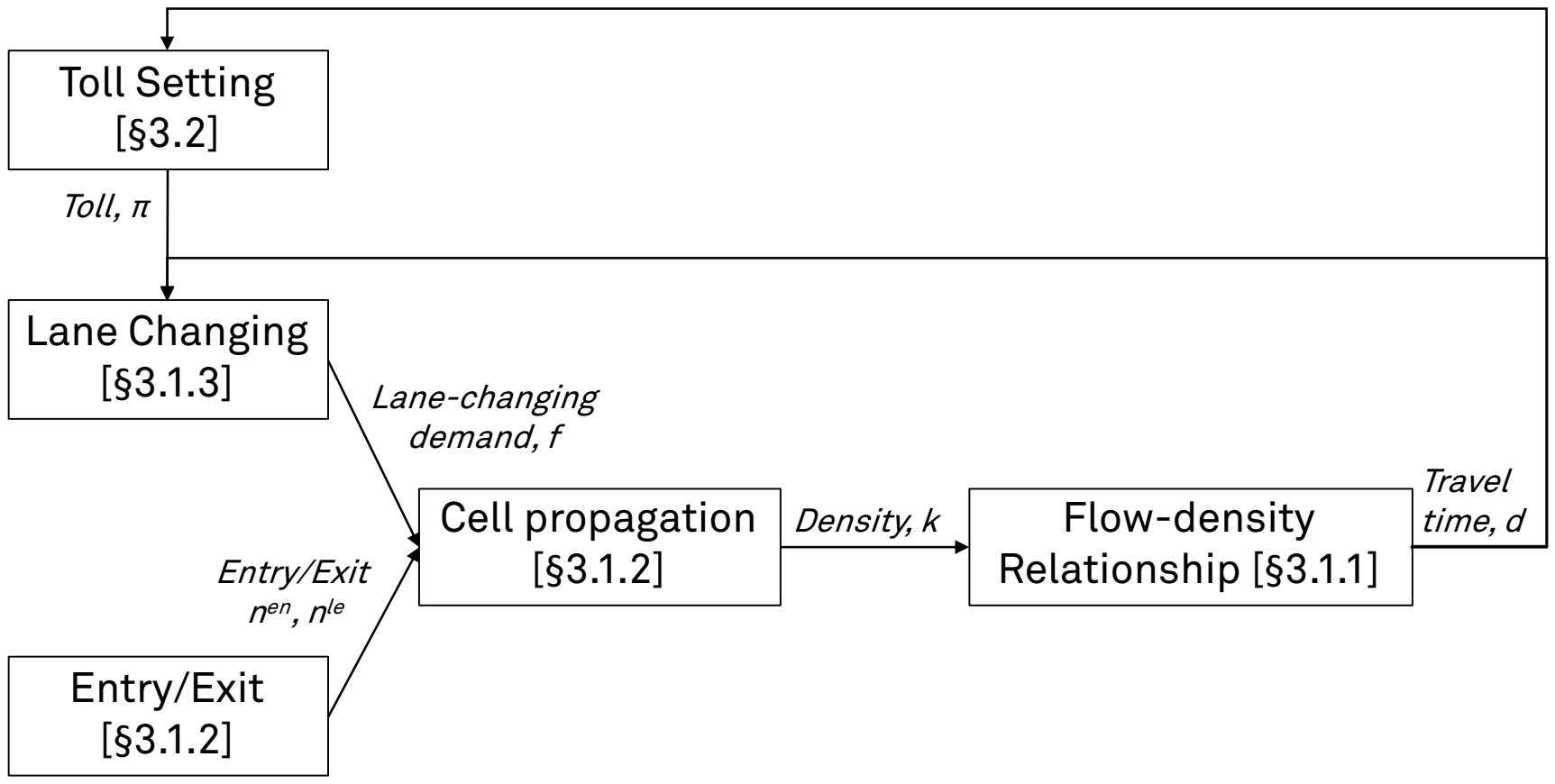

**Figure 1 - Mesoscopic Finite-Difference Traffic Model**

This study focuses on modeling CAVs in MLs and the effects of tolls. Section 3 introduces a finite-difference model (Figure 1) of a multi-lane highway with a single ML (Figure 2). A modified fundamental diagram (FD) by Shi and Li (*5*) incorporates the effects of CAVs on traffic capacities, densities, and speeds. Previous studies on dedicated CAV lanes show that more frequent lane changing will be expected. Therefore, with a similar approach to Roncoli et al. (*6*), the model will explicitly capture lane-changing activities required to utilize MLs along the highway with multiple lane entries/exits. Based on the model, dynamic tolling varying with time and location is adopted, with reactive toll (*7*) set up to avoid flow breakdown. In certain scenarios, high-/low-occupancy CAVs (HOCAVs/LOCAVs) are allowed to use the HOV lanes with or without a toll, which increases the ML traffic capacity. This also reduces GPL flows and therefore alleviates congestion, benefiting low-occupancy HDVs (LOHDVs). A particular study into the effects of HOHDV-HOCAV conversion quantifies such benefits and highlights the policy potential to incentivize such conversion.

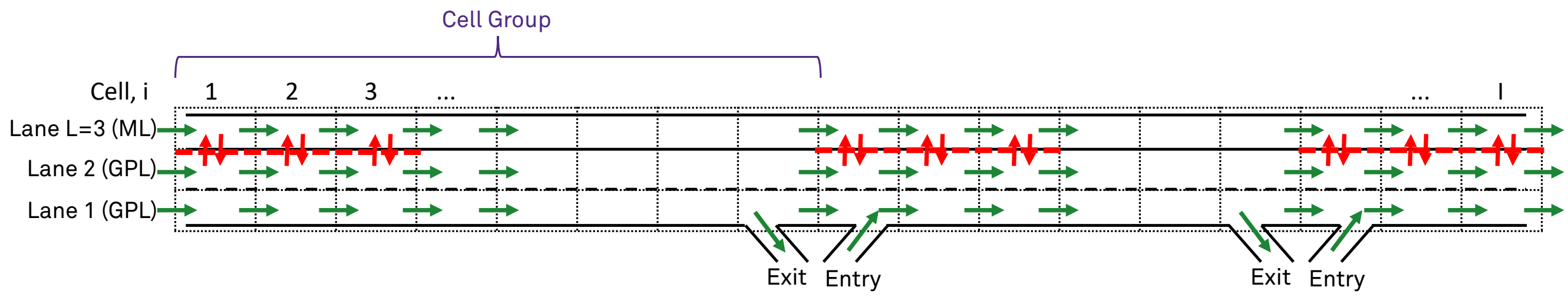

**Figure 2 - Illustration of Highway Managed Lane (ML) Mesoscopic Model**

The mesoscopic model is demonstrated with Monte Carlo simulation in Section 4. Traffic patterns from simulated examples showcase the interactions between toll and traffic, followed by detailed comparisons across various ML policies. Results show that under limited CAV MPRs, free ML usage concentrates CAVs in MLs, attaining the lowest social cost and travel time for all vehicle groups (HOVs, CAVs, and HDVs). Small and targeted tolling avoids flow breakdown in MLs and prioritizes HOVs and other vehicles with high values of time (VOTs), combining the optima of traffic dynamics and allocation economics. Sensitivity analysis with respect to CAV toll levels supports this finding, as lower CAV tolls lead to higher ML capacity. Further sensitivity tests inform the optimal tolling strategy under different CAV MPRs, with particular insights corroborating the proposed HOHDV-HOCAV conversion incentives.

This paper develops a framework for assessing the performance of various ML usage/tolling regimes under mixed traffic flows in different CAV/HOV MPRs, with respect to highway characteristics, traffic pattern, and CAV/HDV interactions. The mesoscopic finite-difference traffic model considers HDV/CAV flow-density relationship, lane-changing behavior, multiple entries/exits, and a reactive toll-setting mechanism. Results show that combined usage of CAVs/HOVs with a targeted and small HDV toll can avoid flow breakdown and minimize total social cost at limited CAV MPRs, achieving synergy between traffic science concepts of flow stability and the economics of resource allocation. Benefits also arise from converting high-occupancy HDVs to CAVs, leading to potential policy incentives.

## 2 LITERATURE REVIEW

This section first reviews previous research efforts in traditional congestion pricing and lane-based tolls, followed by the more recent research on CAVs in mixed traffic flow, which introduces new challenges in traffic modeling and opportunities for new toll regimes.

### 2.1 Conventional Congestion Pricing

In the past half century, numerous studies, e.g., (*8–11*) have looked into congestion pricing, often in an attempt to solve the morning commute problem with bottlenecks from both economic and traffic perspectives, with the objective of attaining better system efficiency by nudging users from a user equilibrium towards system optimum.

Recent literature has covered dynamic tolls for heterogeneous users. Lu et al. (*12*) proposed a bi-criterion dynamic user equilibrium (BDUE) model in a network for time-varying toll charges on heterogeneous users with different VOT preferences. Two approaches to take into account user heterogeneity were discussed - discretizing feasible VOT range into intervals and assuming a continuous VOT distribution.

Furthermore, Wardrop's first principle was extended as "*For each OD pair and for each departure time interval, every trip-maker cannot decrease the experienced generalized trip cost with respect to that trip's particular VOT α by unilaterally changing paths*". The problem was then formulated as infinite-dimensional variational inequality and solved by a simulation-based heuristic. The algorithm to seek a time-dependent shortest path, i.e., with the least generalized cost, interacts with the mesoscopic traffic simulator for the flow physics (DYNASMART) and updates the path set and assignment using a gap-based approach (or alternatively the simpler but less efficient method of successive averages) (*13*).

Dong and Mahmassani (*14*) further expanded the concept of toll to incorporate the probability of flow breakdown. Noting the Weibull distribution of breakdown probability with respect to pre-breakdown flow rates, the paper formulates the generalized cost as expected travel times weighted by the probability of flow breakdown. A reliability toll is then charged based on the predicted flow rates and the probability over the prediction horizon. Research has also looked into other approaches to incorporate traffic physics into the problem. Simoni et al. (*15*) added spatial distribution to the Network Fundamental Diagram and evaluated two kinds of step tolls with flat tolls. Agent-based simulation with MATSim is carried out on the Zurich network.

## 2.2 Managed Lane (ML) / High Occupancy Toll (HOT) Lane

The idea of Lu et al. (*12*) was extended to reactive and anticipatory dynamic pricing by Dong et al. (*16*). Two problems were identified in the pre-existing dynamic pricing operation, namely late toll adjustment, causing traffic breakdown and expectation deviation of drivers after payment, and flow instabilities, arising from fluctuations in user choice considering differentials of speed and toll between lanes. It was proposed to utilize predicted and real-time traffic conditions to set time-varying tolls to maintain a target level of service (LOS) and avoid traffic breakdowns because of an overdue adjustment of tolls. The maximum allowable flow to maintain the minimum LOS was assigned to MLs, while the others stayed in GPLs in a dynamic second-best price regime.

The reactive strategy (Figure 3) consisted of a closed-loop control mechanism with a toll generator to update the toll based on current link concentration in a linear/non-linear manner. The anticipatory pricing framework (Figure 4) added a traffic prediction component to preemptively update the toll and prevent breakdowns in MLs. The two frameworks were modeled with traditional static tolls in a real-time traffic estimation and prediction system such as DYNASMART-X. A modified Greenshield's model with calibrated dual regimes is used by setting the cut-off concentration corresponding to the target LOS.

The model results showed that predictive price determination played a significant role in achieving the objective with an appropriate price level to maintain target LOS in MLs and no observable deterioration of performance in GPLs.

Lou and Laval (*7*) made use of a multi-lane hybrid traffic flow model by Laval and Daganzo (*17*) to reflect lane-changing behavior on a freeway with one ML and one GPL. It modeled each lane as an individual kinematic wave stream disturbed by vehicles changing lanes and blocking the traffic. The optimization objective was to maximize the throughput with a constraint of ML density not smaller than the critical density. The result showed that lane-changing vehicles behaved like moving bottlenecks. Nevertheless, this behavior probably would not exist in a coordinated CAV system. The paper also discussed the use of loop detector data for recursively learning willingness-to-pay.

Laval et al. (*18*) suggested differentiating pricing for MLs and GPLs to maximize revenue. With a bottleneck model, it was shown that the revenue and delay are functions of pricing, and maximum revenue could be achieved by the highest possible toll while keeping MLs at capacity with no queue.

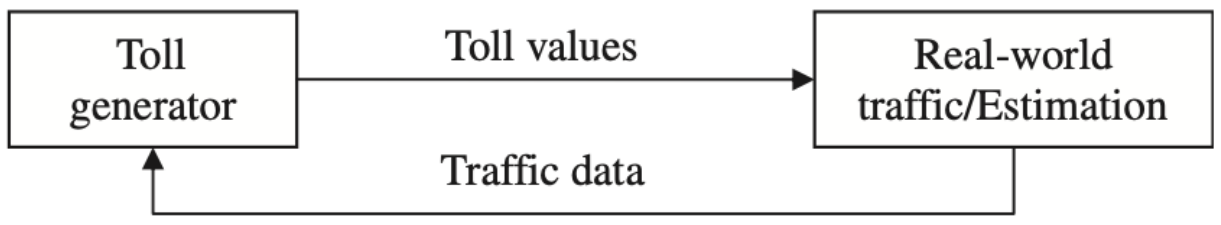

**Figure 3 - Reactive Pricing Framework (*16*)**

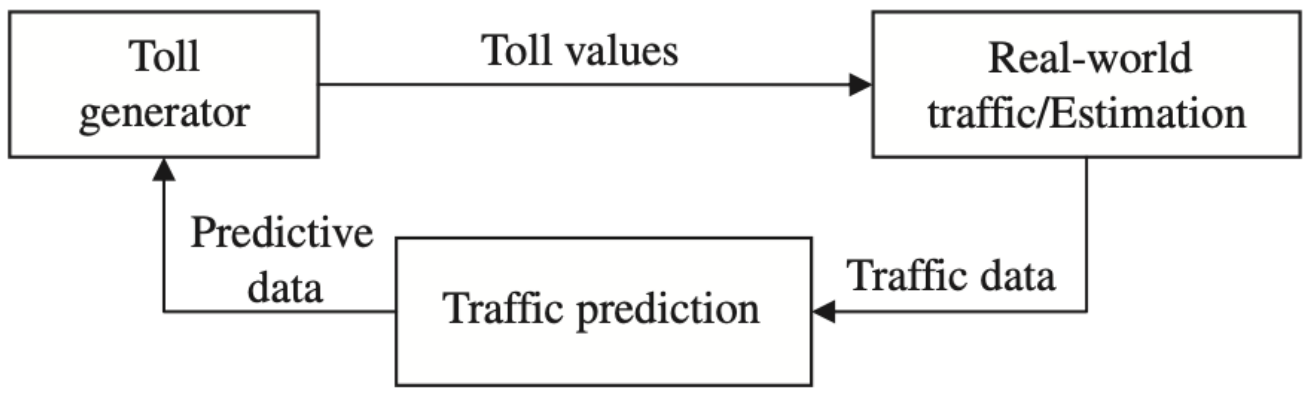

**Figure 4 - Anticipatory Pricing Framework (*16*)**

## 2.3 Connected and Automated Vehicles (CAV)

Recent literature has explored congestion pricing in a CAV network. A review of congestion pricing with CAVs manifested its potential to improve current second-best pricing systems to first-best and enhance the ease of access by handling the computation and communication (*19*). Reinvesting the toll revenue can also lead to public acceptance as an indirect refund. Sharon et al. (*20*) capitalized on the real-time communications capability of CAVs to propose "*Δ-tolling*", which is a continuously updated toll level to bring traffic towards free flow with an exponential decay effect. Branded as model-free without knowledge requirement on traffic demand, the proposed mechanism only needed current travel time and parameters in the tolling formula. Three traffic models were considered in the Sioux Falls and Austin networks. The first model was the traditional link performance function with deterministic and homogeneous travelers. The latter two, cell transmission model and microsimulation, featured heterogeneous travelers and real-time route updates. Significant potential benefits were illustrated in terms of average travel time and total social welfare.

The impact of CAVs on traffic flow at different market penetration flows is a relatively new topic. Talebpour et al. (*2*) suggested dedicated lanes for CAVs only for a relatively high MPR due to congestion and breakdown formation otherwise. Optional use of the lanes was found to yield the greatest benefit. Mohajerpoor and Ramezani (*21*) derived an analytical expression assuming binomial distribution on the number of CAVs in traffic streams and suggested the choice of mixed-HDV, dedicated CAV, and mixed-CAV lanes under different MPRs with validation of microsimulation in Aimsun. Lane-changing is another significant area affecting the traffic flow, in particular in the ML case. Talebpour et al. (*22*, *23*) modeled CAV lane-changing behavior with a game theory approach, as opposed to previous rule-based or gap-acceptance models. Roncoli et al. (*6*) proposed a piecewise linear cell-transmission model to compute lane-changing and longitudinal flow on a multi-lane highway, which Wang et al. (*24*) extended to explore the system-wide impact of ego-efficient CAVs at different MPRs.

Given the varying effect of improving traffic flow with CAV MPRs, researchers looked into combining a CAV lane with an HDV toll lane. Liu and Song (*3*) introduced a problem of autonomous vehicle/toll lanes to cater to the benefits of dedicated lanes for CAVs while accommodating some HDVs willing to pay a toll. However, a modified BPR function with human-driven vehicle equivalent, instead of flow-density relationship, was used to model the impacts of mixed-autonomy traffic on capacities. To determine its optimal toll rate, Wang et al. (*4*) proposed a multi-class traffic assignment problem with elastic demand and consideration of travelers' route choice based on knowledge of traffic conditions under a similar form of a macroscopic simulation and a modified BPR function.

There was recent research considering the utilization of high occupancy lanes by CAVs. The question explored by Guo and Ma (*25*) was the closest to this paper, taking into account mixed-autonomy traffic in MLs. It showed that CAVs could benefit the traffic flow by reducing delay and increasing throughput, mainly through platooning. Nevertheless, the paper did not propose a mechanism to set the toll and investigate its impact on vehicle behavior. Li et al. (*26*) considered assigning toll lanes to CAVs with high occupancy, while allowing other classes of vehicles (human-driven or low occupancy) to pay a toll. A user equilibrium formulation was proposed with the traffic considered by macroscopic link performance functions without specific consideration of CAVs, which limited the model accuracy. Therefore, a research gap remains to investigate the impacts, especially on traffic flow and travel time, of different toll settings and usage policies under the condition of mixed traffic flows with CAVs, HOVs, and HDVs.

The problem characteristics and solution methods of various studies in the literature are summarized in Table 1.

| Reference | Traffic Physics | Scope | Toll Nature | VOT Nature | Solution Method |
|---|---|---|---|---|---|
| Dong et al., 2011 (*16*) | Mesoscopic simulation (DYNASMART-X) with Modified Greenshield's model | GPL/ML | Dynamic | Heterogeneous | Simulation-based heuristic |
| Laval et al., 2015 (*18*) | Bottleneck constraint | GPL/ML | Dynamic | Homogeneous | Numerical analysis |
| Liu & Song, 2019 (*3*) | Modified BPR function for CAVs/HDVs | Network | Static | Homogeneous | Genetic algorithm |
| Lou et al., 2011 (*7*) | Multi-lane hybrid traffic flow model (*17*) | GPL/ML | Dynamic | Homogeneous | Non-linear optimization |
| Lu et al., 2008 (*12*) | Mesoscopic simulation (DYNASMART) | Network | Dynamic | Heterogeneous | Simulation-based heuristic |
| Sharon et al., 2017 (*20*) | a) Link delay function<br>b) Cell transmission model<br>c) Microsimulation | Network with CAVs | Dynamic | a) Homogeneous<br>b & c) Heterogeneous | a) Numerical method<br>b) DTA Simulator (*27*)<br>c) AIM4 microsimulator |
| Simoni et al., 2015 (*15*) | Network Fundamental Diagram with spatial distribution | Network | Dynamic | Homogeneous | Agent-based simulation (MATSim) |
| Wang et al., 2021 (*4*) | Modified BPR function for CAVs/HDVs | Network | Dynamic | Heterogeneous | Norm-relaxed method of feasible direction (NRMFD) |
| This paper | Finite difference model and mixed-traffic flow density for CAVs/HDVs | GPL/ML | Dynamic | Heterogeneous | Simulation-based heuristic |

**Table 1 - Problem Characteristics and Solution Methods of Various Literature**

# 3 MATHEMATICAL FORMULATION

This paper primarily considers reactive pricing in (*7*). As introduced in Section 1, mesoscopic traffic simulation is carried out with a finite difference model, mixed-traffic flow-density relationship, and lane-changing mechanism (Figure 1) on a simplified three-lane highway with one ML as shown in Figure 2.

The assumptions of the problem are as follows:

1. **Fixed demand**: omitting the effects on demand due to overall system performance change (e.g., travel time saving).
2. **Constant but heterogeneous VOTs for individuals**: assuming a linear VOT-time relationship for ease of computation, i.e., $\rho(d) = \alpha d$.
3. **Same physical setting for MLs and GPLs**: but different management regimes.

Eight ML usage and toll-setting policies considered in this paper are summarized in Table 2. The first four policies are exclusive use policies, while EU4 tolls CAVs for using MLs. The last four allow LOHDVs to use MLs with different toll settings.

Parameter notations are summarized in Table 3. Lanes $l < L$ and $l = L$ refer to multiple GPLs and the single ML respectively.

**Table 2 - Policies of Managed Lane (ML) Usage and Toll Setting**

| Policy Group | ML Policy | HOV | LOCAV | LOHDV |
|---|---|---|---|---|
| Exclusive use | EU1. HOV-only | Use | / | / |
| | EU2. CAV-only | / | Use | / |
| | EU3. HOV/CAV-only | Use | Use | / |
| | EU4. HOV/CAV-only & LOCAV-toll | Use | **Toll** | / |
| All use | AU1. All-use | Use | Use | Use |
| Selective toll | ST1. LOHDV-toll | Use | Use | **Toll** |
| | ST2. LOCAV/LOHDV-toll | Use | **Toll** | **Toll** |
| All toll | AT1. All-toll | **Toll** | **Toll** | **Toll** |

**Table 3 - Parameter Notation**

| Category | Notation | Description | Simulation Value (Section 4) |
|---|---|---|---|
| Vehicle | $CAV_k$ | = 1 if vehicle $k$ is a CAV; = 0 if HDV | $\begin{cases} 0: w.p. 0.6 \\ 1: w.p. 0.4 \end{cases}$ |
| | $NP_k$ | Number of passengers in vehicle $k$ | $\begin{cases} 1: w.p. 0.8 \\ 2: w.p. 0.1 \\ 3: w.p. 0.1 \end{cases}$ |
| | $x_{k,i,l}$ | = 1 if vehicle $k$ is in cell $i$ of lane $l$; = 0 otherwise | / |
| | $x_{k,i,l,t}$ | = 1 if vehicle $k$ is in cell $i$ of lane $l$ at time $t$; = 0 otherwise | / |
| | $y^m_{k,i,l,t}$ | = 1 if vehicle $k$ intend to leave cell $i$ at time $t$ for lane $l+m$ or cell $i+1$ if $m=0$; = 0 otherwise | / |

| **Category** | **Notation** | **Description** | **Simulation Value (Section 4)** |
|---|---|---|---|
| | $R_{k,g}^{l}$ | = 1 if vehicle *k* can travel in ML *l* in cell group *g*; = 0 otherwise | / |
| | $start_k$ | Cell group where vehicle *k* enters | $\begin{cases} 0: w.p.\,0.6 \\ 1: w.p.\,0.1 \\ 2: w.p.\,0.1 \\ 3: w.p.\,0.1 \\ 4: w.p.\,0.1 \end{cases}$ |
| | $end_k$ | Cell group where vehicle *k* exits | $\begin{cases} 0: w.p.\,0.05 \\ 1: w.p.\,0.05 \\ 2: w.p.\,0.05 \\ 3: w.p.\,0.05 \\ 4: w.p.\,0.8 \end{cases}$ |
| Cost | $\alpha_k$ | VOT of vehicle *k* (USD/h) | $(\sim N(20,10) \in [0.5,300]) \times NP_k$ |
| | $C_k$ | Total cost of vehicle *k* | / |
| | $\pi_k$ | Total toll charged on vehicle *k* | / |
| | $\pi_{glh}$ | Toll charged on vehicles in cell group *g* at time horizon *h* in lane *l* (per cell) | / |
| | $\pi_{min}$ | Minimum toll level (USD) | 0 |
| | $\pi_{max}$ | Maximum toll level (USD) | 15 |
| | $\pi_{max,k}$ | Maximum toll level for vehicle *k*, considering the CAV/HOV/HDV status and toll policy (USD) | / |
| | $\pi_{step}$ | Toll adjustment step (USD) | 0.2 |
| | $TT$ | Total toll | / |
| | $GC_{g,l,t,k}$ | Generalized cost of traveling in cell group *g* in lane *l* at time *t* for vehicle *k* | / |
| | $TC$ | Total social cost (total user cost minus total toll) | / |
| | $TC_{i,t}$ | Total social cost of cell *i* at time *t* | / |
| | $\delta_{LC}$ | Nominal lane changing cost (USD) | 0.1 |
| Time | $d_k$ | Total travel time of vehicle *k* | / |
| | $d_{i,l,t}$ | Travel time in cell *i* of lane *l* at time *t* | / |
| | $d(q_H, q_A)$ | Travel time function with HDV flow $q_H$ and CAV flow $q_A$ for one cell | / |
| | $d_f$ | Free flow travel time | / |
| | $t$ | Index of time | / |
| | $\Delta t$ | Time increment (s) | 6 |
| | $t^*{}_k$ | Departure time of vehicle *k* (h, am) | ~trapezoidal (7,7.5,8.5,9) |
| | $t_a$ | Model start time (h, am) | 7 |
| | $t_b$ | Model end time (h, am) | 10 |

| **Category** | **Notation** | **Description** | **Simulation Value (Section 4)** |
|---|---|---|---|
| | $h$ | Index of time period | / |
| | $\Delta \text{h}$ | Time period of considering previous travel time for toll setting (min) | 5 |
| | $t(h)$ | Start time of time period *h* | / |
| | $h(t)$ | Time period of time *t* | / |
| | $T$ | Number of time periods considered | / |
| | $TTT$ | Total travel time | / |
| Traffic | $n_{i,l,t}^{0}$ | Number of vehicles flowing from cell *i* to cell *i*+*1* in lane *l* at time *t* | / |
| | $n_{i,l,t}^{m}$ | Number of vehicles in cell *i* flowing from lane *l* to lane *l*+*m* at time *t;* $m = \{1, -1\}$ | / |
| | $n_{i,0,t}^{en}$ | Number of vehicles entering the highway in cell *i* | / |
| | $n_{i,0,t}^{le}$ | Number of vehicles leaving the highway in cell *i* | / |
| | $q_{i,l,t}$ | Flow of vehicles in cell *i* of lane *l* at time *t* | / |
| | $f_{i,l,t}^{m}$ | Demand of vehicles in cell *i* of lane *l* at time *t* to change lanes to *l*+*m;* $m = \{1,0,-1\}$ | / |
| | $n_{i,l,t}$ | Number of vehicles in cell *i* of lane *l* at time *t* | / |
| | $n_{H,i,l,t}$ | Number of HDVs in cell *i* of lane *l* at time *t* | / |
| | $n_{A,i,l,t}$ | Number of CAVs in cell *i* of lane *l* at time *t* | / |
| | $q_{i,l,t}^{max}$ | Maximum flow of cell *i* of lane *l* at time *t* | / |
| | $q_{H}^{m}$ | Capacity of HDV-only traffic flow (veh/h) | 1800 |
| | $q_{A}^{m}$ | Capacity of CAV-only traffic flow (veh/h) | 2600 |
| | $q_{H}^{0}$ | Intercept of HDV congested-regime branch and flow rate axis in HDV-only traffic flow FD (veh/h) | 2424 |
| | $q_{A}^{0}$ | Intercept of CAV congested-regime branch and flow rate axis in CAV-only traffic flow FD (veh/h) | 4400 |
| | $w_H$ | Backward shockwave speed of HDV-only traffic flow (km/h) | 30.5 |
| | $w_A$ | Backward shockwave speed of CAV-only traffic flow (km/h) | 61.1 |
| | $k_{j,H}$ | Jam density of HDV-only traffic flow (veh/km) | 94.4 |
| | $k_{j,A}$ | Jam density of CAV-only traffic flow (veh/km) | 75.0 |
| | $k_{i,l,t}$ | Density in cell *i* of lane *l* at time *t* (veh/km) | / |
| | $k_{i,l,t}^{cr}$ | Critical density where traffic transitions from free flow to congestion in cell *i* of lane *l* at time *t* (veh/km) | / |
| | $s_f$ | Free flow speed (km/h) | 88 |
| | $s_0$ | Minimum speed (km/h) | 5 |
| | $\theta_{toll}$ | Proportion of critical density which would trigger toll change | 0.85 |
| Model | $l$ | Index of lane | / |

| Category | Notation | Description | Simulation Value (Section 4) |
|---|---|---|---|
| | $L$ | Number of lanes | 3 |
| | $L_t$ | Total length of highway (km) | 10 |
| | $L_i$ | Length of cell *i* (km) | 0.133 |
| | $i$ | Index of cells | / |
| | $N_i$ | Number of cells | 75 |
| | $N_{lc}$ | Number of cells for ML lane changing in a cell group | 3 |
| | $g$ | Index of cell groups | / |
| | $g(i)$ | Cell group index for cell *i* | / |
| | $GI(g)$ | Set of cells in cell group *g* | / |
| | $N_g$ | Number of cell groups | 5 |
| | $N_k$ | Number of vehicles in Monte Carlo simulation | 6000 |
| | $N_{iter}$ | Number of iterations in Monte Carlo simulation | 100 |

## 3.1 Traffic Simulation Model

### *3.1.1 Flow-density Relationship*

The relationship between flow, $q_{i,l,t}$, and speed of cell *i* in lane *l* at time *t*, is estimated in Eq. (1) with the mixed traffic flow FD proposed by Shi and Li (*5*) to incorporate the effect of CAVs on traffic flow.

$$q_{i,l,t} = \begin{cases} v_f k_{i,l,t} & : k_{i,l,t} \in \left[0, k_{i,l,t}^{cr}\right] \\ \dfrac{n_{i,l,t} - k_{i,l,t}\left(\dfrac{w_H n_{H,i,l,t}}{q_H^0} + \dfrac{w_A n_{A,i,l,t}}{q_A^0}\right)}{\dfrac{n_{H,i,l,t}}{q_H^0} + \dfrac{n_{A,i,l,t}}{q_A^0}} & : k_{i,l,t} \in \left[k_{i,l,t}^{cr}, \dfrac{n_{i,l,t}}{\dfrac{w_H n_{H,i,l,t}}{q_H^0} + \dfrac{w_A n_{A,i,l,t}}{q_A^0}}\right] \end{cases} \quad ,\forall i,l,t \tag{1}$$

The critical density, $k_{i,l,t}^{cr}$, is influenced by the proportion of CAVs in each cell in Eq. (2):

$$k_{i,l,t}^{cr} = \frac{n_{i,l,t}}{\dfrac{(s_f + w_H) n_{H,i,l,t}}{q_H^0} + \dfrac{(s_f + w_A) n_{A,i,l,t}}{q_A^0}}, \forall i,l,t \tag{2}$$

The maximum flow, $q_{i,l,t}^{max}$, can then be obtained by Eq. (3):

$$q_{i,l,t}^{max} = k_{i,l,t}^{cr} s_f, \forall i,l,t \tag{3}$$

### *3.1.2 Finite Difference Model*

To assess the possibility of shifting vehicles from GPLs to the ML, each lane of the freeway is discretized into cells (Figure 2). Entries to and exits from the highway are allowed at the start of cell groups, as shown by the bottom diagonal green arrows.

Lane changing between the ML and GPLs are shown as red dotted lines, i.e., vehicles can only enter/leave the ML at designated cells, while vehicles can change lanes freely in GPLs. Vehicles are modeled with heterogeneous VOTs, and would choose to enter the ML if the policy allows and the VOT saving is higher than the toll. Cell groups are defined to be the set of consecutive cells at the start of ML lane change to the next start of ML lane change. This serves as the basis of travel time evaluation, toll setting, and drivers' decisions to change lanes. Lane changing happens across cells, instead of at the end of cells. Priority is given to through traffic, then fast lane to slow lane, before slow lane to fast lane.

Eq. (4) stipulates the number of vehicles that can flow to the next cell, based on the minimum of demand ($n_{i,l,t}$ - the number of vehicles in the cell) and supply ($q^0_{i,l,t}\Delta t$ - maximum cell transfer capacity and $\left(k^{cr}_{i,l,t} - k_{i+1,l,t}\right)\mathrm{L_i}$ - the number of available gaps in the next cell):

$$n^0_{i,l,t} = \min\left\{n_{i,l,t}, q^0_{i,l,t}\Delta t, \left(k^{cr}_{i,l,t} - k_{i+1,l,t}\right)\mathrm{L_i}\right\}, \forall i,l,t \quad (4)$$

Similar to Roncoli et al. (*6*), Eq. (5) calculates the number of vehicles that can change lanes ($n^m_{i,l,t}$, where $m = 1$ to faster lane or $m = -1$ to slower lane) based on the demand, $f^m_{i,l,t}$ (further discussed in Section 3.1.3), and supply, $q^0_{i,l,t}\Delta t\left(1 - k_{i,l+m,t}/k^{cr}_{i,l,t}\right)$, the latter of which is proportional to the number of available gaps in the desired lane:

$$n^m_{i,l,t} = \min\left\{f^m_{i,l,t}, q^0_{i,l,t}\Delta t\left(1 - k_{i,l+m,t}/k^{cr}_{i,l,t}\right)\right\}, \forall i,l,t,m \in \{-1,1\} \quad (5)$$

Eq. (6) updates the cell occupancies, $n_{i,l,t+1}$, at the next time period based on the inflow, outflow, lane changing, and entry/exit:

$$n_{i,l,t+1} = n_{i,l,t} - n^0_{i,l,t} - n^1_{i,l,t} - n^{-1}_{i,l,t} + n^0_{i-1,l,t} + n^1_{i,l-1,t} + n^{-1}_{i,l+1,t} + n^{en}_{i,0,t} - n^{le}_{i,0,t}, \forall i,l,t \quad (6)$$

Eq. (7)-(8) calculate the density, $k_{i,l,t}$, and flow, $q_{i,l,t}$,:

$$k_{i,l,t} = \frac{n_{i,l,t}}{L_i}, \forall i,l,t \quad (7)$$

$$q_{i,l,t} = n_{i,l,t}/d_{l,i,t}, \forall i,l,t \quad (8)$$

### *3.1.3 Lane Changing Decisions*

Similar to Roncoli et al. (*6*), several assumptions on lane-changing behavior apply for realism:

1. Vehicles request to change lanes if:
    a. They need to do so (e.g., to exit the highway); or
    b. The road configuration allows (e.g., CAV-only ML), and changing lanes will bring a lower perceived generalized cost, $GC_{g,l,t,k}$ of cell group *g* in lane *l* at time *t* for vehicle *k*, which is the sum of expected travel cost and toll in Equation (9).
2. Vehicles in GPLs give up changing to the faster lane if no gap is available.
3. Vehicles which request to change to a slower lane continue to move forward if there is no gap in the next lane, unless that is the last cell for the vehicles to change lanes. In that case, the vehicles forcefully change to the slower lane, possibly causing overcrowding in the slower lane.

$$GC_{g,l,t,k} = \sum_{i \in GI(g)} \left(\pi_{glt} + \alpha_k d_{i,l,t}\right), \forall g, l, t, k \tag{9}$$

The restrictions are implemented through variables $R_{k,g}^{L}$, which is set to 1 if the vehicle *k* can travel in MLs considering the toll policy and whether it enters/exits in cell group *g* (i.e., vehicles must enter/exit the highway in the slowest lane, except when they start with the highway.) (Equation (10)):

$$R_{k,g}^{L} = \begin{cases} 1 \ \mathit{if}\ start_k < g < end_k \ \mathit{and} \begin{cases} NP_k > 1\ \mathit{for}\ EU1 \\ CAV_k = 1\ \mathit{for}\ EU2 \\ NP_k > 1\ \mathit{or}\ CAV_k = 1\ \mathit{for}\ EU3, EU4 \\ AU1, ST1, ST2, AT1 \end{cases} \\ 0\ \mathit{otherwise} \end{cases}, \forall k, g \tag{10}$$

The lane-switching criteria are then formulated in Eq. (11)-(12). The drivers only change lanes ($y_{k,i,l,t}^{-1}$ or $y_{k,i,l,t}^{1} = 1$) if the perceived benefits are greater than the threshold $\delta_{LC}$, and only change to the lane with lower generalized cost.

$$y_{k,i,l,t}^{-1} = \begin{cases} 1\ \mathit{if}\ R_{k,g(i)+1}^{l} = 0\ \mathit{or}\ (GC_{g(i)+1,l-1,t,k} + \delta_{LC} < GC_{g(i)+1,l,t,k} \\ \quad \mathit{and}\ GC_{g(i)+1,l-1,t,k} < GC_{g(i)+1,l+1,t,k}) \\ 0\ \mathit{otherwise} \end{cases}, \forall k, i, l, t \tag{11}$$

$$y_{k,i,l,t}^{1} = \begin{cases} 1\ \mathit{if}\ R_{k,g(i)+1}^{l+1} = 1\ \mathit{and}\ (GC_{g(i)+1,l+1,t,k} + \delta_{LC} < GC_{g(i)+1,l,t,k} \\ \quad \mathit{and}\ GC_{g(i)+1,l+1,t,k} < GC_{g(i)+1,l-1,t,k}) \\ 0\ \mathit{otherwise} \end{cases}, \forall k, i, l, t \tag{12}$$

Summing the individual decisions would form the total lane changing demand, $f_{i,l,t}^{m}$, in Eq. (13), which will then be fed into Equation (5) in Section 3.1.2 for determining whether the lane changing is successful for each vehicle:

$$f_{i,l,t}^{m} = \sum_{k} y_{k,i,l,t}^{m}, \forall i, l, t, m \in \{-1,1\} \tag{13}$$

### 3.2 Toll Setting Mechanism

Toll level is set to keep the ML traffic stable at high capacity without traffic breakdown. The reactive tolling mechanism adjusts the toll level, $\pi_{glh}$ for cell group *g* in lane *l* at time horizon *h*, according to the current traffic in MLs in Eq. (14) to avoid exceeding a pre-set percentage (e.g., 85%) of the critical density between free-flow and congestion regimes. The toll, capped by minimum and maximum limits, is applied to the whole cell group. A time horizon, *h*, is considered to avoid excessive fluctuation of tolls. Drivers "pay" the toll after they pass the lane changing cells.

$$\pi_{glh} = \begin{cases} \min\left(\pi_{max}, \pi_{g,l,h-1} + \pi_{step}\right) if \sum_{i \in GI(g)} \sum_{t=t(h-1)}^{t(h)-1} k_{i,l,t} \geq \theta_{toll} \sum_{i \in GI(g)} \sum_{t=t(h-1)}^{t(h)-1} k_{i,l,t}^{cr} \\ \max\left(\pi_{min}, \pi_{g,l,h-1} - \pi_{step}\right) if \sum_{i \in GI(g)} \sum_{t=t(h-1)}^{t(h)-1} k_{i,l,t} < \theta_{toll} \sum_{i \in GI(g)} \sum_{t=t(h-1)}^{t(h)-1} k_{i,l,t}^{cr} \end{cases} \quad (14)$$

$$, \forall g, h, l = N_l - 1$$

The toll, $\pi_k$, total toll, $TT$, travel time, $d_k$, total travel time, $TTT$, and total cost, $TC$, of vehicle $k$ are then evaluated in Eq. (15)-(19), noting that $\pi_{max,k}$ stipulates the maximum toll a vehicle should pay based on its CAV/HOV/HDV status and prevailing toll policy:

$$\pi_k = \sum_g \sum_l \sum_t \left( \min\left(\pi_{g(i)lh(t)}\, x_{k,\left(\frac{N_i}{N_g}\right)g+N_{lc},l,t}, \pi_{max,k}\right)\right), \forall k \quad (15)$$

$$TT = \sum_t \sum_i \sum_l \pi_{g(i),l,h(t)}\, n_{i,l,t} = \sum_k \pi_k \quad (16)$$

$$d_k = \sum_t \sum_i \sum_l \left(d_{i,l,t} x_{k,i,l,t}\right), \forall k \quad (17)$$

$$TTT = \sum_t \sum_i \sum_l d_{i,l,t}\, n_{i,l,t} = \sum_k d_k \quad (18)$$

$$TC = \sum_k \alpha_k d_k \quad (19)$$

### 3.3 Effect of Human-Driven Vehicle (HDV) - Connected and Automated Vehicle (CAV) Conversion on Traffic

An interesting question arising from the above derivation is the effect of CAV MPR on traffic. Based on the FD of (*5*) in Equations (1) and (2), more CAVs in a lane would increase its capacity and allow vehicles from GPLs to enter. This would reduce the journey time of those vehicles, as well as others remaining in GPLs due to relieved congestion.

The net benefits brought by the conversion of HOHDVs to HOCAVs can be estimated with the assumption that the free flow in the ML is maintained, the traffic in GPL is kept at the congestion regime, and the additional capacity in the ML is filled up by vehicles switching from GPLs. For simplicity, the VOT is taken as the mean value. The additional benefits brought by the improved performance in lane changing by CAVs are also ignored.

The marginal cost brought by converting an HOV to a CAV in ML, $\frac{\Delta TC_{i,t}}{\Delta n_{A,i,l,t}}$, is estimated in Equation (20), where the first coefficient part accounts for the number of additional vehicles that can move from the GPL $l'$ to the ML due to the increased capacity, as converted from the density derivative, $\frac{\partial k_{i,l,t}^{cr}}{\partial n_{A,i,l,t}}$ in Equation (21), which is based on the FD in Equation (2). The coefficient is multiplied by two due to the net effect between adding one CAV (higher critical density) and one HDV (lower critical density). Then the first term after that is the VOT gained by each vehicle moving from the GPL to the ML. The second term is the VOT saved by vehicles remaining in the GPL due to relieved congestion, as calculated from the derivative of travel time against the number of vehicles, $\frac{\partial d_{i,l,t}}{\partial n_{i,l,t}}$ in Equation (22). The travel time derivative is again calculated with the FD in Equation (1) by conservatively assuming only HDVs are present, which underestimates the travel time saved given the higher CAV backward shockwave speed.

$$\frac{\Delta TC_{i,t}}{\Delta n_{A,i,l,t}} \approx -\left[2\frac{\partial k_{i,l,t}^{cr}}{\partial n_{A,i,l,t}} L_i\right]\left[\left(d_{i,l,t} - d_{i,l',t}\right)\bar{\alpha} + \frac{\partial d_{i,l',t}}{\partial n_{i,l',t}} q_{i,l',t} \frac{L_i}{s_f} \bar{\alpha}\right] \quad ,\forall i,t,l = L > l' \tag{20}$$

$$\frac{\partial k_{i,l,t}^{cr}}{\partial n_{A,i,l,t}} = \frac{\frac{\left(s_f + w_H\right)n_{H,i,l,t}}{q_H^0} - \frac{\left(s_f + w_A\right)n_{H,i,l,t}}{q_A^0}}{\left(\frac{\left(s_f + w_H\right)n_{H,i,l,t}}{q_H^0} + \frac{\left(s_f + w_A\right)n_{A,i,l,t}}{q_A^0}\right)^2}, \forall i,t,l = L \tag{21}$$

$$\frac{\partial d_{i,l,t}}{\partial n_{i,l,t}} = \frac{1}{q_{i,l,t}} - L\left(\frac{k_{i,l,t}}{{q_{i,l,t}}^2}\right)\left(\frac{\partial q_{i,l,t}}{\partial n_{i,l,t}}\right) = \frac{1}{q_{i,l,t}} + \frac{k_{i,l,t} w_H}{{q_{i,l,t}}^2}, \forall i,t,l < L \tag{22}$$

This benefit is particularly relevant if we convert vehicles originally in ML for free, i.e., HOVs. It may be worth incentivizing the conversion of HOHDVs to HOCAVs due to the positive externalities brought to other vehicles and the whole network. The formulae enable numerical evaluation of benefits in converting one HOHDV to a HOCAV. Assuming the values in Table 3, with a 40% CAV MPR in the ML, a near-capacity ML ($k_L = 0.85k^{cr}$) and a saturated GPL ($k_{l'} = 63veh/km$), $\frac{\partial k_{i,l,t}^{cr}}{\partial n_{A,i,l,t}} \approx 1.86veh/km/veh$, $\frac{\partial d_{i,l,t}}{\partial n_{i,l,t}} \approx 0.0097\ h/veh$, VOT savings to the additional vehicles shifted to the ML and vehicles remaining in GPLs are \$9.3 and \$13.8 respectively. Considering round-trip and 250 days/year, we attain \$11550 annual benefit, which can constitute a sizeable subsidy to CAV conversion. Section 4.5 continues the discussion with simulation results.

# 4 SIMULATION AND RESULTS

Monte Carlo simulation is carried out to demonstrate the proposed framework, incorporating the finite-difference traffic model and dynamic toll setting mechanism introduced in Figure 1 and Section 3. The ML policy characteristics are shown in Table 2 and the values of parameters are summarized in Table 3.

The model generates 6000 vehicles with a triangular distribution over two hours, equivalent to a peak hour flow of 4000veh/h or 1333veh/h/lane. 60% of vehicles start with the modeled highway, while 10% enter at each entrance. 5% of vehicles leave at each exit, while 80% end with the highway. The CAV MPR is set as 40%. 20% of vehicles are HOVs, half of which have two passengers, and the remaining half have three. Some vehicle parameters are randomly distributed, including CAV capability ($CAV_k$), number of passengers ($NP_k$), VOTs ($\alpha_k$), and departure times ($t^*{}_k$). VOTs are assumed to follow a normal distribution ($N(20{,}10)$) in (*12*) with a bound of [0.5,300] (in USD/hour).

The model is implemented with Python 3.8.10 and run on a computer with an Intel Xeon 3GHz CPU and 128GB memory.

This section continues with the traffic pattern under the optimal ML Policy ST1 (HOV/CAV use, HDV toll), followed by a comparison across policies. Sensitivity analysis of CAV toll levels focuses on the gap between Policies ST1 and ST2, followed by CAV MPRs in various policies to demonstrate the performance variation. Lastly, the HOHDV-HOCDV conversion is evaluated with various HOCAV MPRs. Error bars in plots show standard deviations.

## 4.1 Traffic Pattern – ML Policy ST1 (HOV/CAV use, HDV toll)

Figure 5 shows a sample heat map of flow densities in cells in each lane with time. The backward shock waves caused by entrance/exit effects can be observed at around cell 45 (4th group entrance/exit lane), where the darker cell colors indicate congestion propagating upstream along time. The ML is also observed to have lower traffic density in general.

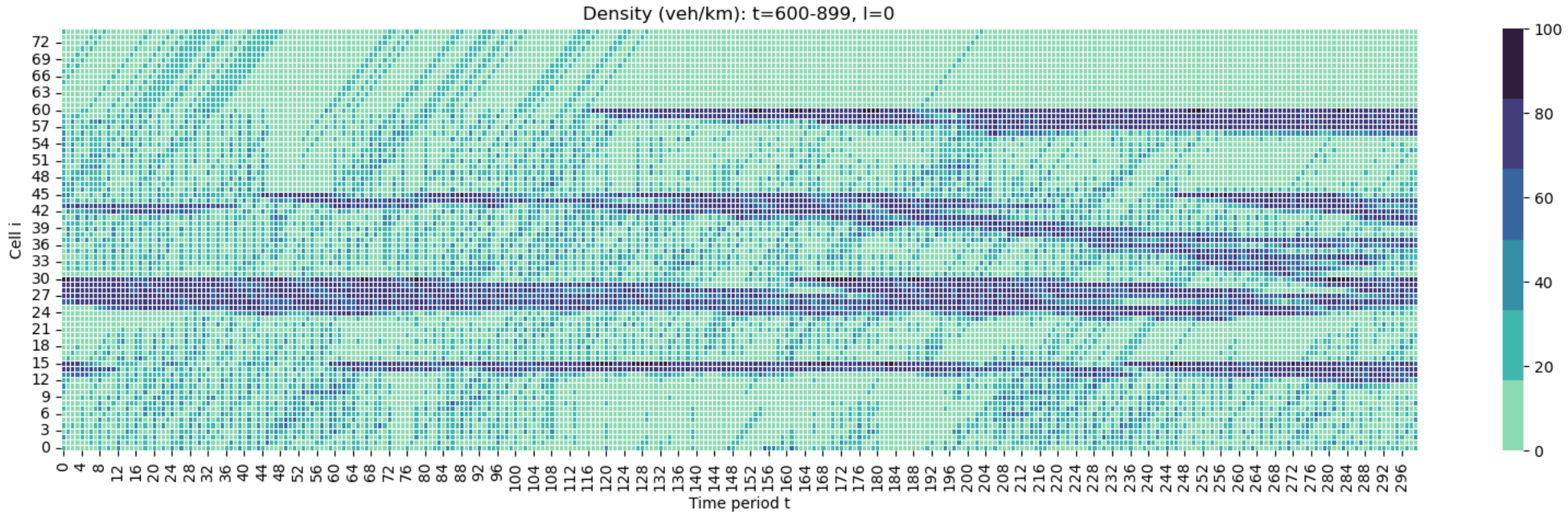

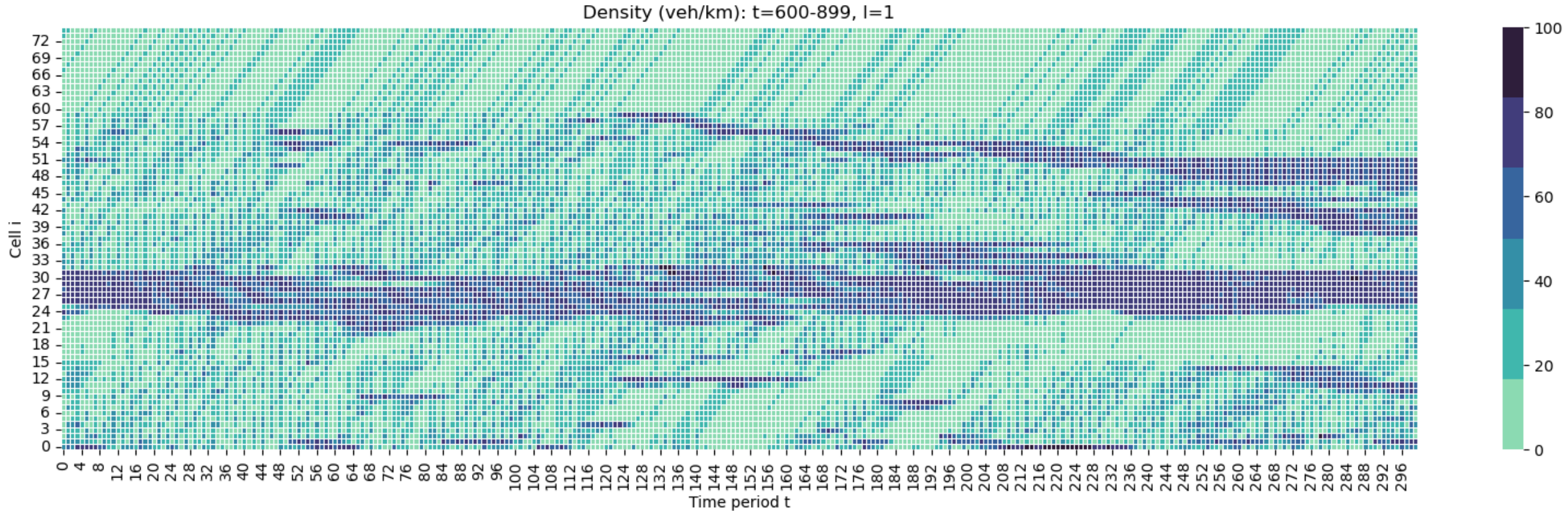

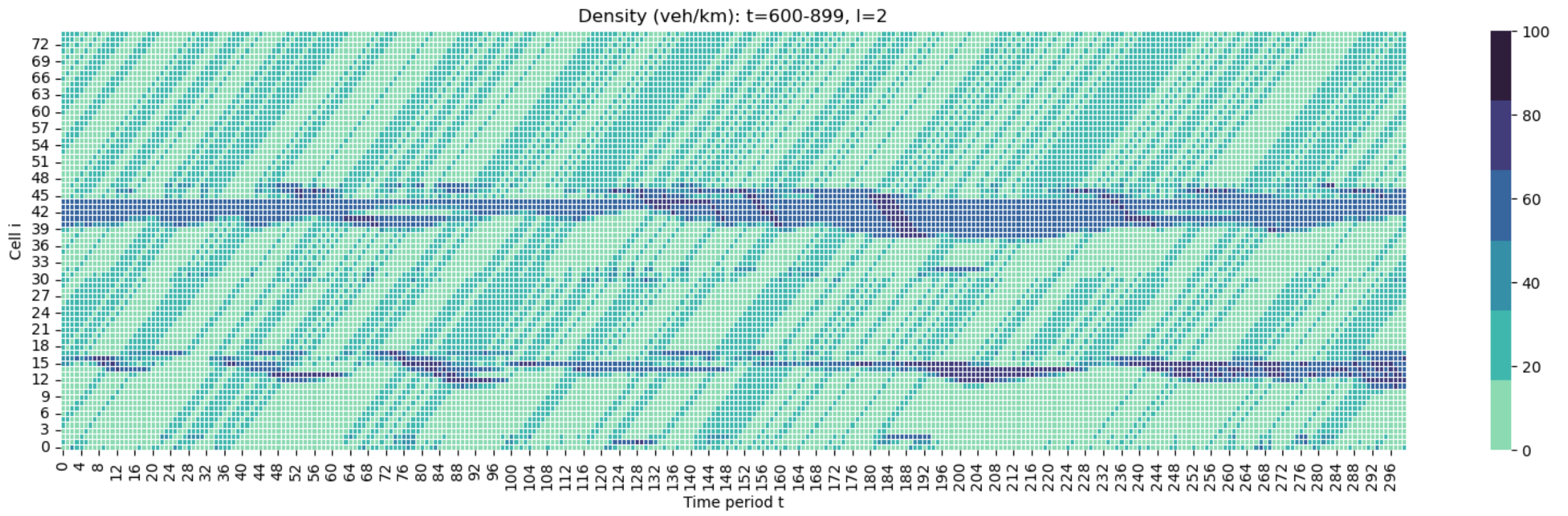

**Figure 5 - Heat Maps of Densities in Lanes – ML Policy ST1**

Figure 6 shows a sample graph of trajectories of every ten vehicles for clarity. The colors of the lines illustrate the types of vehicles, while the types of dotted lines indicate the use of different lanes. The lane changing and overtaking, as well as the difference in lane traveling speed, can be observed from the trajectories.

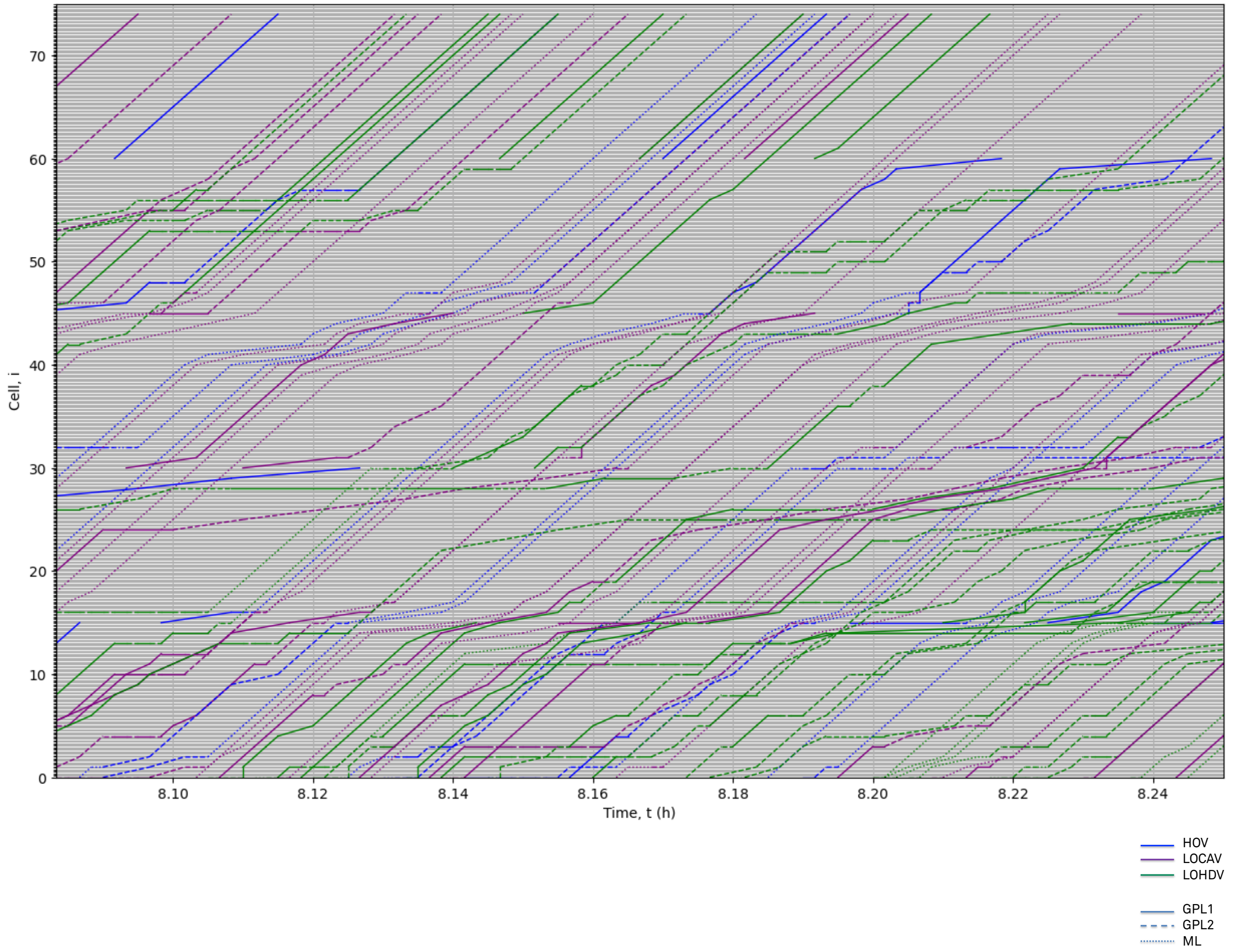

**Figure 6 - Trajectories of Vehicles (every ten vehicles) - ML Policy ST1**

## 4.2 Comparison across Managed Lane (ML) Policies

Figure 7 and Figure 8 summarize the total social cost and average travel time with error bars indicating standard deviation.

ML Policy ST1 (HOV/CAV not tolled; LOHDV tolled) attains the best results in terms of total social costs and average travel time, followed by LOCAV/LOHDV-toll ST2, all-use AU1, and all-toll AT1. This suggests that indiscriminate tolling policies may lead to sub-optimal allocation of traffic capacities and therefore higher social costs.

Among other policies, the all-use policy AU1 shows higher variance probably due to the greater chance of congestion in the ML. All exclusive-use policies incur higher costs and travel times because of excessive congestion in the two GPLs, which also affect HOVs/CAVs to enter/exit the highway, showing that a CAV MPR at 40% requires shared use with LOHDVs to optimize resources.

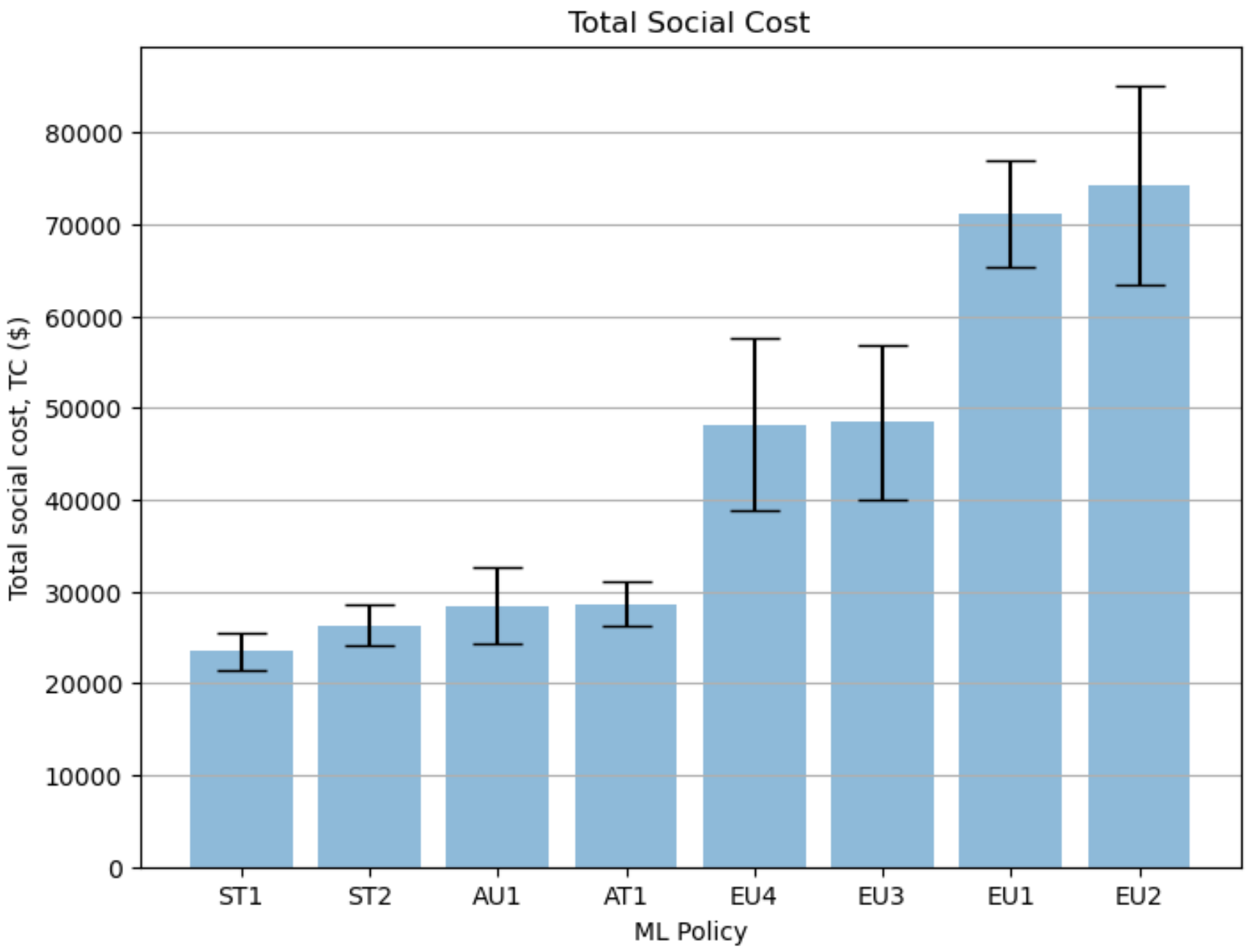

| ML Policy | HOV | LOCAV | LOHDV |
|---|---|---|---|
| EU1 | Use | / | / |
| EU2 | / | Use | / |
| EU3 | Use | Use | / |
| EU4 | Use | **Toll** | / |
| AU1 | Use | Use | Use |
| ST1 | Use | Use | **Toll** |
| ST2 | Use | **Toll** | **Toll** |
| AT1 | **Toll** | **Toll** | **Toll** |

**Figure 7 - Total Social Cost across Managed Lane (ML) Policies**

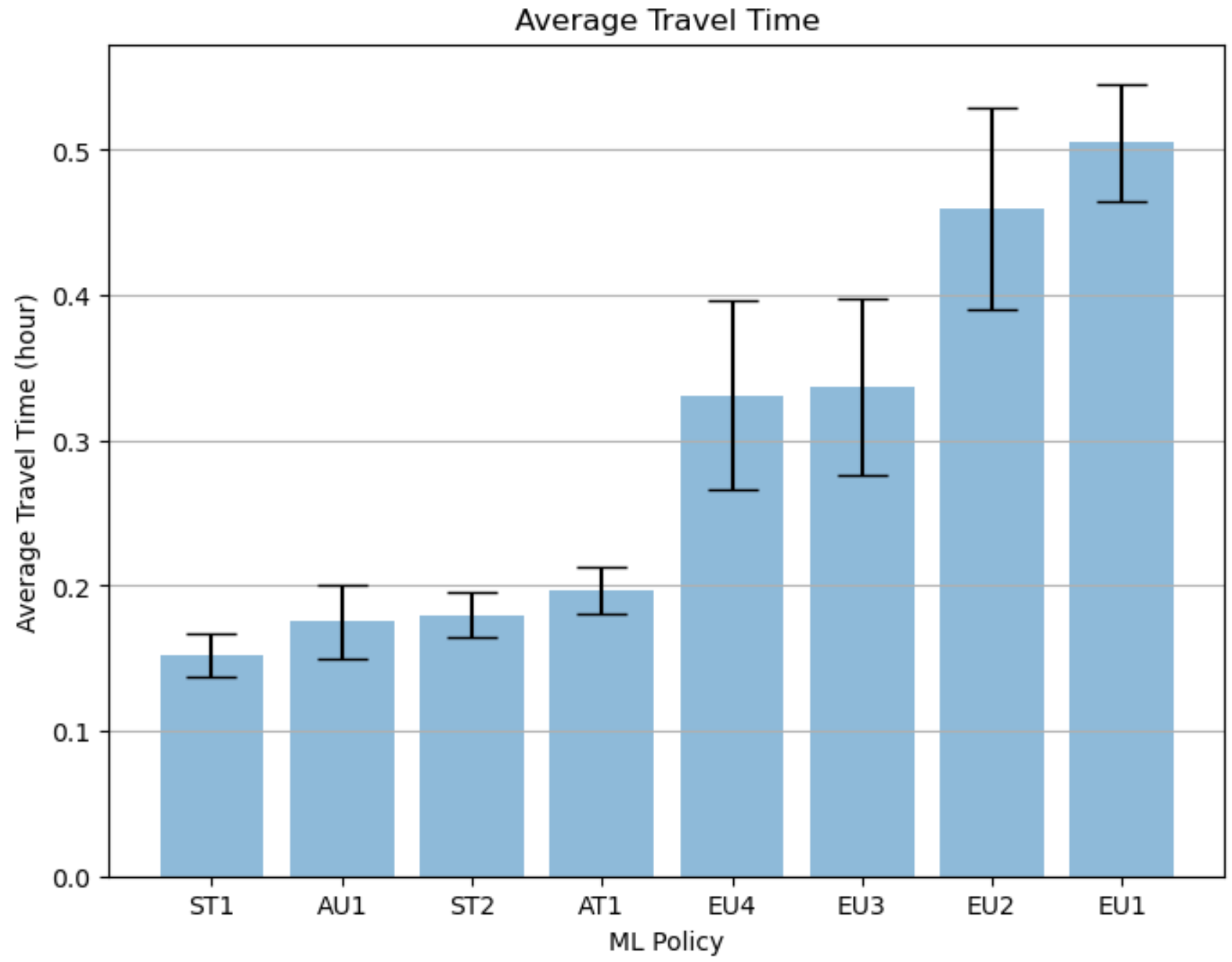

| ML Policy | HOV | LOCAV | LOHDV |
|---|---|---|---|
| EU1 | Use | / | / |
| EU2 | / | Use | / |
| EU3 | Use | Use | / |
| EU4 | Use | **Toll** | / |
| AU1 | Use | Use | Use |
| ST1 | Use | Use | **Toll** |
| ST2 | Use | **Toll** | **Toll** |
| AT1 | **Toll** | **Toll** | **Toll** |

**Figure 8 - Average Travel Time across Managed Lane (ML) Policies**

To examine the results for the four leading policies more closely, Figure 9 and Figure 10 show the average travel time and generalized costs (median of iterations) of vehicles departing at different times. ST1 offers the lowest travel time for overall vehicles and CAVs, but margins shrink in HOVs (vs. ST2) and HDVs (vs. AU1). This suggests ST1 benefits CAVs/HOVs with minimal adverse effects on HDVs. No toll on LOCAVs encourages their concentration in the ML, smoothens traffic flow, and reduces the likelihood of flow breakdown, which will be further discussed in Section 4.3. Meanwhile, the LOHDV toll still allows LOHOVs with high VOTs to take advantage of the low travel time in the ML. ST1 strikes a balance between the traffic science of stable flows and the economics of resource allocation.

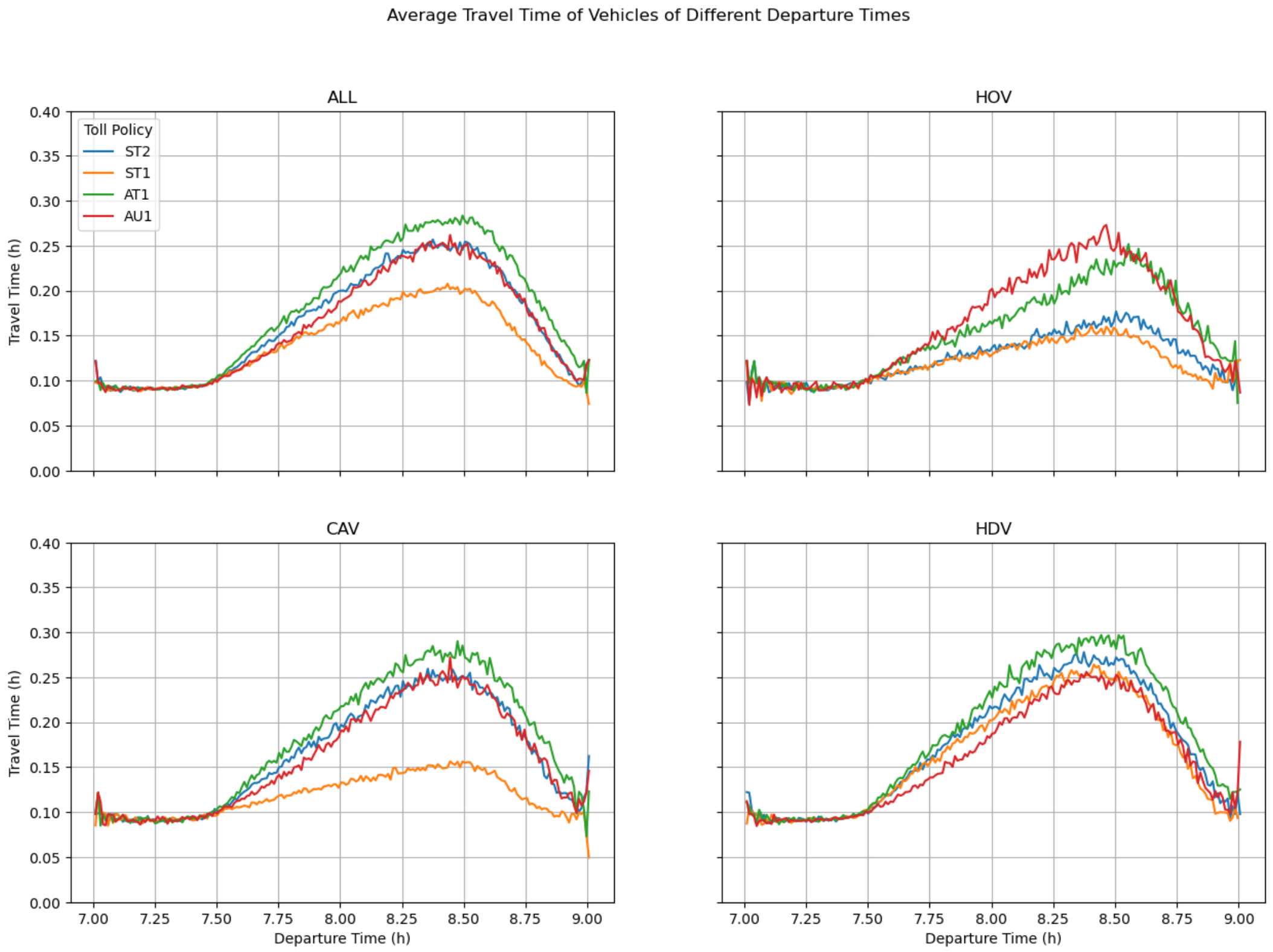

**Figure 9 - Average Travel Time for Different Departure Times across Managed Lane (ML) Policies**

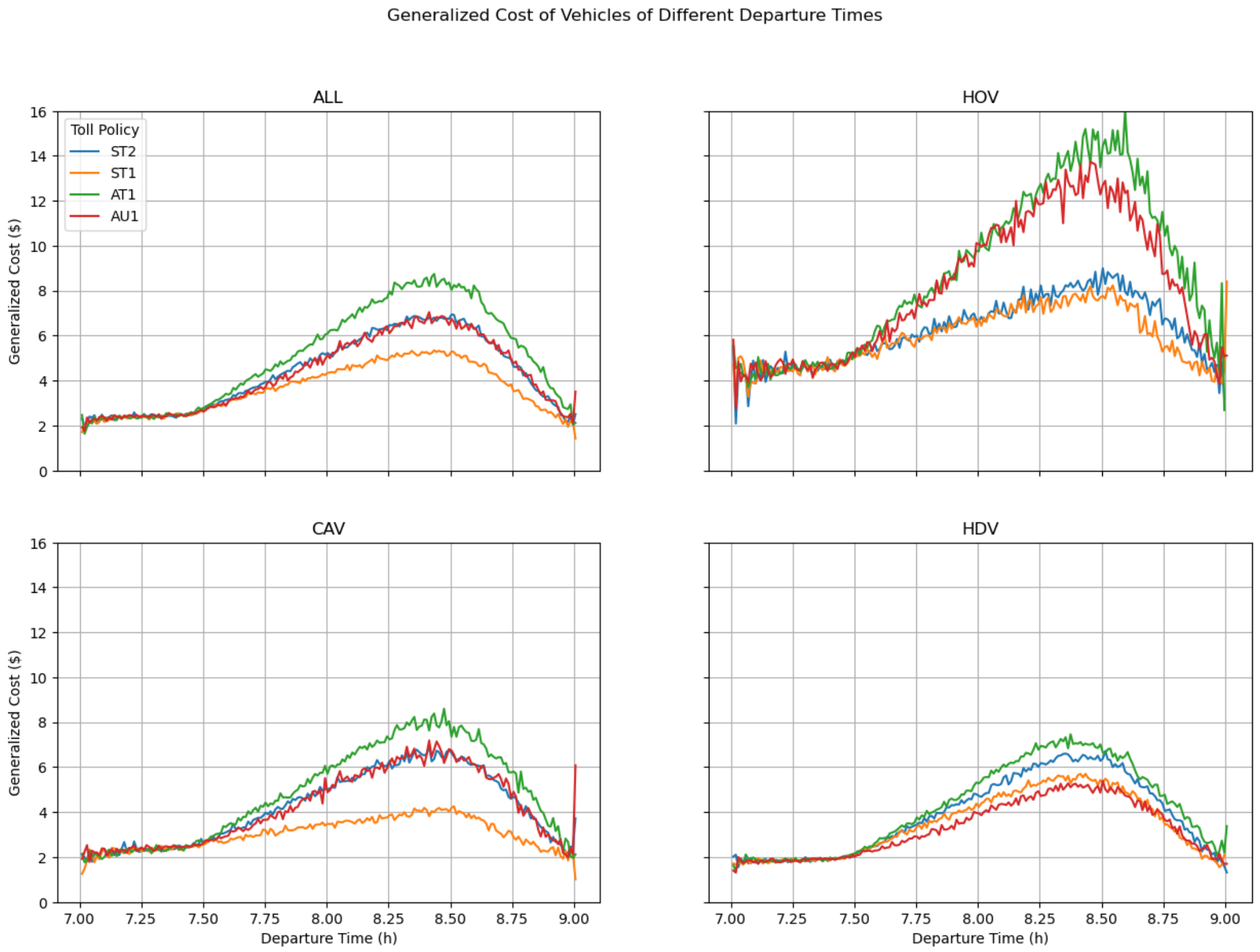

**Figure 10 - Generalized Costs of Different Departure Times across Managed Lane (ML) Policies**

HOVs generally have lower travel time but higher generalized costs due to their higher VOTs. They are willing to pay higher tolls in exchange for faster travel time under ST2 (as shown in Figure 11). CAVs also benefit from lower travel time than HDVs in ST1, but the effect is less obvious in the other three policies, as more vehicles dilute the relative advantage of the ML.

The average toll collected is below $3, comparable to saved VOTs. In particular, on average, ST1 tolls less than $1.5 on HDVs, showing that small and targeted tolls are effective in capturing higher-VOT drivers while avoiding overcrowding the ML.

The detailed results are shown in Table 4.

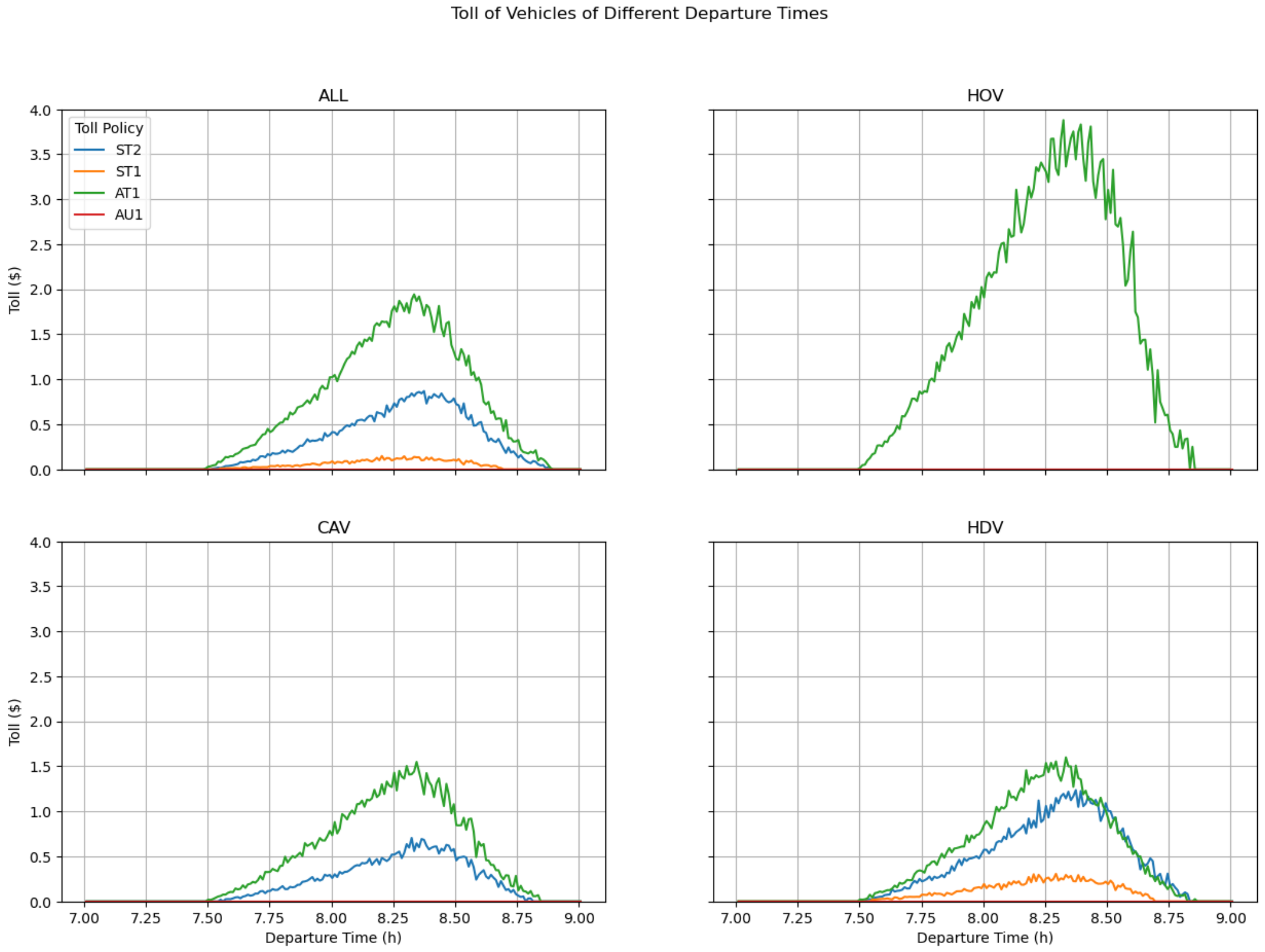

**Figure 11 - Toll on Vehicles across Managed Lane (ML) Policies**

**Table 4 - Results of Comparison across Managed Lane (ML) Policies (95% confidence intervals in brackets)**

| --- ALL VEHICLES --- | | | | | | | | |
|---|---|---|---|---|---|---|---|---|
| **ML Policy** | **EU1** | **EU2** | **EU3** | **EU4** | **ST2** | **ST1** | **AT1** | **AU1** |
| Total toll ($) | / | / | / | 0 (0-28.948) | 2271.2 (1187.2-3803.1) | 463.5 (121.06-1190.1) | 5243.2 (2197.1-7536.3) | / |
| Total tolled car | / | / | / | 0 (0-50.4) | 1111.5 (840.12-1419.6) | 313 (153.5-493.9) | 1875 (1271-2197.1) | / |
| Total tollable car | / | / | / | 1876 (1803.3-1948.6) | 4685.5 (4604.3-4746.5) | 2799 (2725.4-2885.6) | 5858.5 (5832.9-5877.1) | / |
| Avg toll per tolled car ($) | / | / | / | / | 2.0349 (1.15-3.0906) | 1.4178 (0.7216-3.0524) | 2.8107 (1.498-3.6621) | / |
| Tolled % | / | / | / | 0 (0-2.7275) | 23.792 (18.117-30.252) | 11.101 (5.5038-17.412) | 32.003 (21.703-37.446) | / |
| Total vehicle travel time (h) | 3030.7 (2524.5-3436.6) | 2759.7 (2156.8-3577.7) | 2020 (1320.6-2677.3) | 1986.6 (1364.1-2867.2) | 1079.7 (911.58-1259.9) | **912.62 (768.24-1093)** | 1178.4 (1029.8-1368.9) | 1052.7 (846.74-1414.7) |
| Avg travel time (h) | 0.50512 (0.42074-0.57277) | 0.45996 (0.35947-0.59629) | 0.33667 (0.2201-0.44622) | 0.3311 (0.22734-0.47787) | 0.17996 (0.15193-0.20998) | **0.1521 (0.12804-0.18217)** | 0.1964 (0.17163-0.22815) | 0.17546 (0.14112-0.23579) |
| Total drivers' cost ($) | 71205 (59885-81514) | 74221 (58518-94555) | 48421 (32173-63597) | 48180 (33833-70333) | 28552 (24119-33100) | **23893 (20261-28549)** | 33723 (28720-39185) | 28458 (23008-38203) |
| Total social cost ($) | 71205 (59885-81514) | 74221 (58518-94555) | 48421 (32173-63597) | 48178 (33831-70333) | 26316 (22508-31150) | **23480 (20109-27611)** | 28657 (24929-33921) | 28458 (23008-38203) |

| --- Connected and Automated Vehicles (CAVs) --- | | | | | | | | |
|---|---|---|---|---|---|---|---|---|
| **ML Policy** | **EU1** | **EU2** | **EU3** | **EU4** | **ST2** | **ST1** | **AT1** | **AU1** |
| Total toll ($) | / | / | / | 0 (0-28.948) | 741.86 (342.14-1200.4) | / | 1641.8 (653.67-2419.5) | / |
| Total tolled car | / | / | / | 0 (0-50.4) | 447.5 (336.23-572.92) | / | 750 (495.32-884.25) | / |
| Total tollable car | / | / | / | 1876 (1803.3-1948.6) | 1876 (1809.5-1930) | / | 2347 (2262.4-2427.1) | / |
| Avg toll per tolled car ($) | / | / | / | / | 1.6322 (0.87262-2.3751) | / | 2.175 (1.1581-2.896) | / |
| Tolled % | / | / | / | 0 (0-2.7275) | 23.961 (17.654-30.459) | / | 32.265 (20.939-37.704) | / |
| Total vehicle travel time (h) | 1214.6 (1017-1380.6) | 646.35 (487.28-847.61) | 516.25 (369.74-672.41) | 519.9 (383.6-719.78) | 433.67 (364.24-505.15) | **299.05 (266.86-346.3)** | 470.41 (410.56-548.15) | 419.3 (343.35-570.62) |
| Avg travel time (h) | 0.50471 (0.4225-0.57033) | 0.27158 (0.20601-0.35559) | 0.21447 (0.15202-0.28149) | 0.21554 (0.15977-0.30174) | 0.18039 (0.15115-0.21081) | **0.12446 (0.11228-0.14307)** | 0.19556 (0.17093-0.22755) | 0.17534 (0.14132-0.23927) |
| Total drivers' cost ($) | 28582 (23433-33092) | 17009 (12765-22255) | 13383 (9388.9-17595) | 13574 (10182-19382) | 11230 (9510.3-13007) | **8048 (7147.7-9333.9)** | 13125 (11174-15083) | 11343 (9305.9-15274) |
| Total social cost ($) | 28582 (23433-33092) | 17009 (12765-22255) | 13383 (9388.9-17595) | 13563 (10180-19382) | 10470 (8915.7-12415) | **8048 (7147.7-9333.9)** | 11554 (9828-13297) | 11343 (9305.9-15274) |

| --- High-occupancy Vehicles (HOVs) --- | | | | | | | | |
|---|---|---|---|---|---|---|---|---|
| **ML Policy** | **EU1** | **EU2** | **EU3** | **EU4** | **ST2** | **ST1** | **AT1** | **AU1** |
| Total toll ($) | / | / | / | / | / | / | 2149.7 (915.85-3084.3) | / |
| Total tolled car | / | / | / | / | / | / | 577 (436.65-670.2) | / |
| Total tollable car | / | / | / | / | / | / | 1167.5 (1117.5-1226.7) | / |
| Avg toll per tolled car ($) | / | / | / | / | / | / | 3.7714 (2.0099-4.8525) | / |
| Tolled % | / | / | / | / | / | / | 49.271 (36.846-56.679) | / |
| Total vehicle travel time (h) | 287.46 (242.45-341.81) | 552.27 (427.71-708.23) | 245.45 (181.44-313.95) | 247.51 (181.49-355.07) | 159.47 (137-192.64) | **150.22 (133.69-173.17)** | 195.67 (166.25-238.16) | 212.52 (172.33-291.17) |
| Avg travel time (h) | 0.24021 (0.19997-0.28599) | 0.46135 (0.36073-0.60144) | 0.2061 (0.1462-0.26185) | 0.20525 (0.15506-0.28475) | 0.13253 (0.11569-0.16235) | **0.12462 (0.11308-0.14279)** | 0.16303 (0.14205-0.19435) | 0.17836 (0.14405-0.23959) |
| Total drivers' cost ($) | 14632 (12376-17932) | 28523 (22130-36173) | 12446 (9265.8-16102) | 12569 (9350.4-18541) | 8191.6 (7112.7-9949.2) | **7710.3 (6900.6-8961.5)** | 11770 (9789.8-13680) | 11020 (8883.9-14920) |
| Total social cost ($) | 14632 (12376-17932) | 28523 (22130-36173) | 12446 (9265.8-16102) | 12569 (9350.4-18541) | 8191.6 (7112.7-9949.2) | **7710.3 (6900.6-8961.5)** | 9521.8 (7945.9-11773) | 11020 (8883.9-14920) |

| **--- Human-driven Vehicles (HDVs) ---** | | | | | | | | |
|---|---|---|---|---|---|---|---|---|
| **ML Policy** | **EU1** | **EU2** | **EU3** | **EU4** | **ST2** | **ST1** | **AT1** | **AU1** |
| Total toll ($) | / | / | / | / | 1521.8 (815.92-2623.9) | 463.5 (121.06-1190.1) | 2068.8 (856.44-3087.8) | / |
| Total tolled car | / | / | / | / | 669 (502.48-845.05) | 313 (153.5-493.9) | 760.5 (516.7-953.67) | / |
| Total tollable car | / | / | / | / | 2807 (2731.3-2870.8) | 2799 (2725.4-2885.6) | 2811.5 (2732.9-2911) | / |
| Avg toll per tolled car ($) | / | / | / | / | 2.3015 (1.3435-3.5884) | 1.4178 (0.7216-3.0524) | 2.7029 (1.4567-3.7195) | / |
| Tolled % | / | / | / | / | 23.926 (18.246-29.876) | 11.101 (5.5038-17.412) | 27.092 (18.382-33.734) | / |
| Total vehicle travel time (h) | 1628.3 (1339-1859.6) | 1696.5 (1295.1-2207.2) | 1345.7 (837.49-1833.9) | 1321.3 (874.64-1925.4) | 550.76 (462.74-643.35) | 522.4 (406.27-641.68) | 586.88 (508.88-693.98) | **502.15 (399.84-680.06)** |
| Avg travel time (h) | 0.56628 (0.47508-0.64447) | 0.58737 (0.4537-0.75212) | 0.47056 (0.29392-0.62371) | 0.4575 (0.30569-0.6724) | 0.19179 (0.16157-0.22567) | 0.1833 (0.14367-0.22323) | 0.20439 (0.17647-0.23736) | **0.17432 (0.14096-0.23415)** |
| Total drivers' cost ($) | 33470 (27301-38376) | 34939 (26932-45223) | 27701 (17165-37876) | 27531 (18058-39597) | 12534 (10286-14832) | 11203 (8558.2-13877) | 13547 (11401-16175) | **10350 (8370.6-14043)** |
| Total social cost ($) | 33470 (27301-38376) | 34939 (26932-45223) | 27701 (17165-37876) | 27531 (18058-39597) | 10892 (9125.6-12795) | 10770 (8418.9-13057) | 11424 (9968.2-13626) | **10350 (8370.6-14043)** |

### 4.3 Sensitivity Analysis: Market Penetration Rate (MPR) of Connected and Autonomous Vehicles (CAVs)

This section examines how the ST1 policy's relative advantage varies under different CAV MPRs. Besides, from Section 4.2, exclusive lane use policies do not perform well at the 40% CAV MPR, but the question still holds at higher CAV MPRs.

From Figure 12 and Figure 13, ST1 outperforms other policies at most CAV MPRs, except ST2 at 10%. This may be explained by the dilemma of introducing CAVs to the ML at a low MPRs, where the improvement in capacity is minimal and offset by the increase in traffic density. In this particular case, a LOCAV toll avoids overcrowding the ML and priorities vehicles with higher VOTs.

Exclusive lane use policies suffer from overcrowding until reaching higher CAV MPRs. EU3 and EU4 bring the lowest social cost and average travel time before converging to AU1, showing no negative effects of LOCAV tolls. The near convergence to non-exclusive tolling policies after 70% also suggests that specific tolling policies may only be relevant during the transition state from HDVs to CAVs.

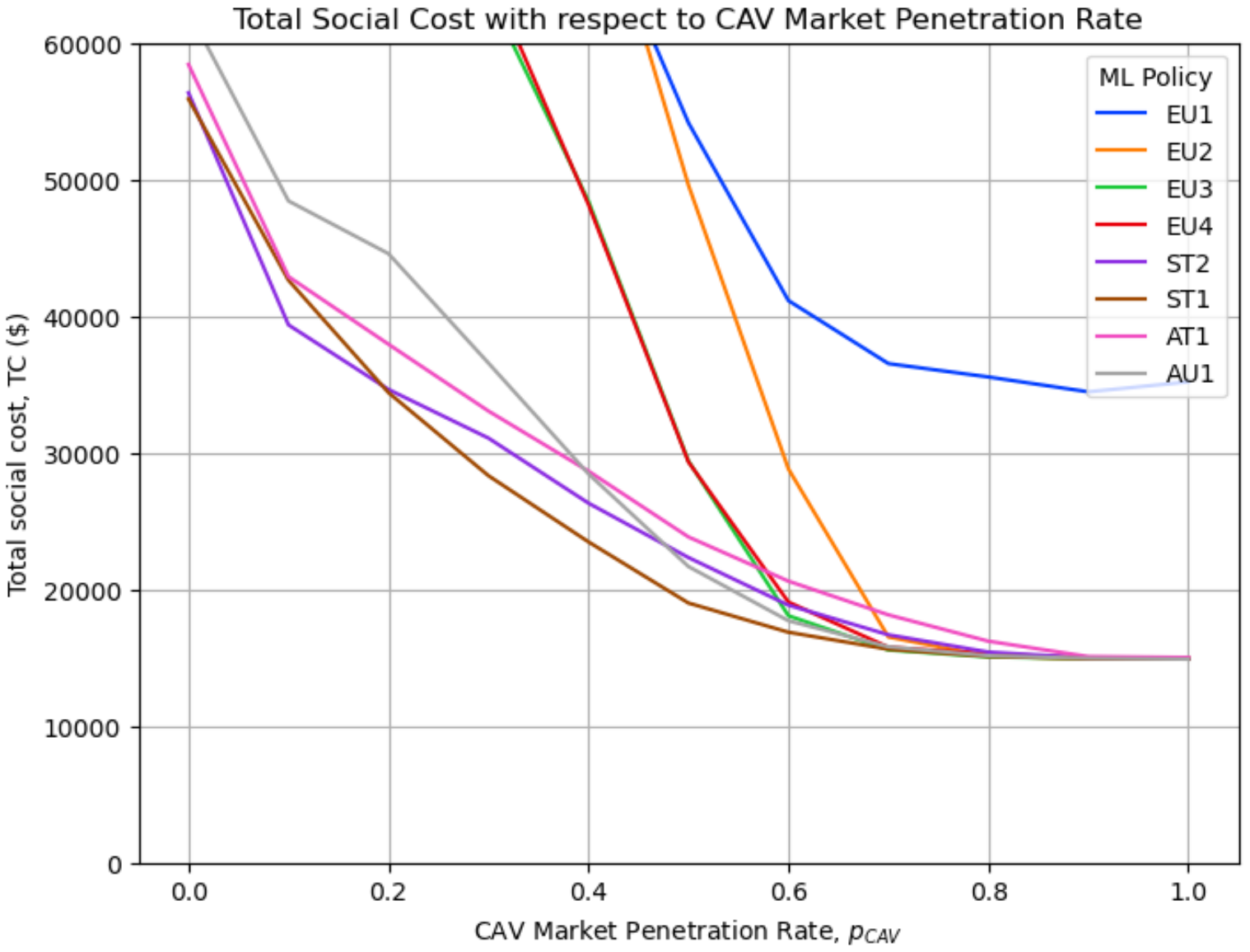

| ML Policy | HOV | LOCAV | LOHDV |
|---|---|---|---|
| EU1 | Use | / | / |
| EU2 | / | Use | / |
| EU3 | Use | Use | / |
| EU4 | Use | **Toll** | / |
| AU1 | Use | Use | Use |
| ST1 | Use | Use | **Toll** |
| ST2 | Use | **Toll** | **Toll** |
| AT1 | **Toll** | **Toll** | **Toll** |

**Figure 12 - Total Social Cost against Connected and Automated Vehicle (CAV) Market Penetration Rate (MPR)**

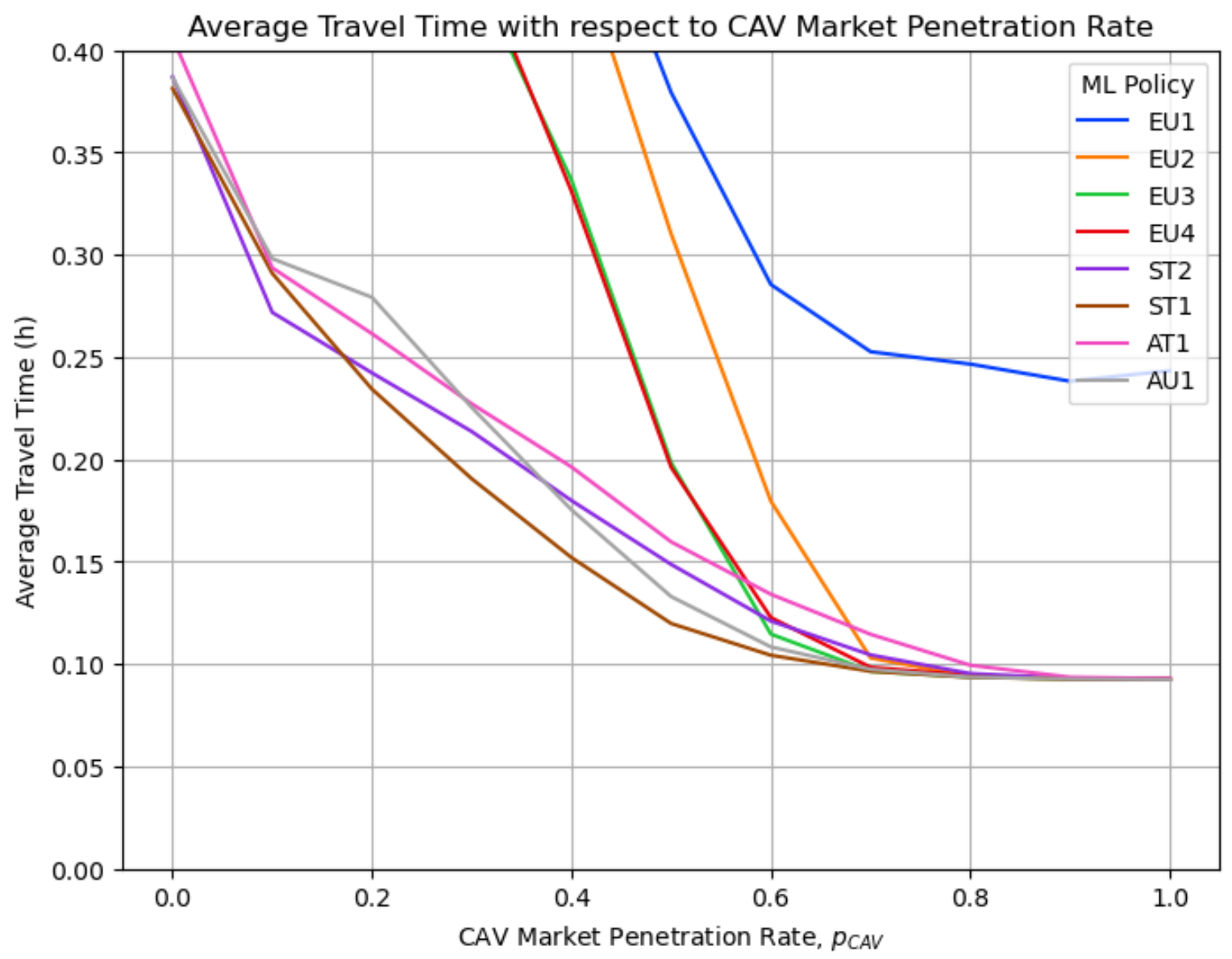

| ML Policy | HOV | LOCAV | LOHDV |
|---|---|---|---|
| EU1 | Use | / | / |
| EU2 | / | Use | / |
| EU3 | Use | Use | / |
| EU4 | Use | **Toll** | / |
| AU1 | Use | Use | Use |
| ST1 | Use | Use | **Toll** |
| ST2 | Use | **Toll** | **Toll** |
| AT1 | **Toll** | **Toll** | **Toll** |

**Figure 13 - Average Travel Time against Connected and Automated Vehicle (CAV) Market Penetration Rate (MPR)**

### 4.4 Sensitivity Analysis: Low-Occupancy Connected and Automated Vehicle (LOCAV) Toll - ML Policy ST2 (HOV/CAV use, Low-Occupancy toll)

Following the result comparison between ML Policies ST1 and ST2 in Section 4.2, sensitivity analysis with different levels of LOCAV tolls under ST2 was carried out to assess further whether a zero LOCAV toll provides the optimal results. In other words, this is to evaluate the gap between the 0% LOCAV toll (of LOHDV) in ST1 and the 100% toll in ST2.

Figure 14 and Figure 15 illustrate a general increasing trend of total social cost and average travel time over increasing the LOCAV toll. This can be explained by the decreasing CAV concentration in the ML shown in Figure 16 (from 75% to 30%). The resulting lower capacity limits more vehicles to use the ML, leading to lower density (Figure 17) and relatively unstable traffic (Figure 18: instantaneous travel time refers to the total travel time on all cells in a lane at a time). This, in turn, raises GPL traffic density (Figure 17) and travel time (Figure 18), even though their CAV proportions are higher (Figure 16).

Therefore, the optimal policy is to impose no toll on CAVs to allow concentration effect, given sufficient CAV MPRs.

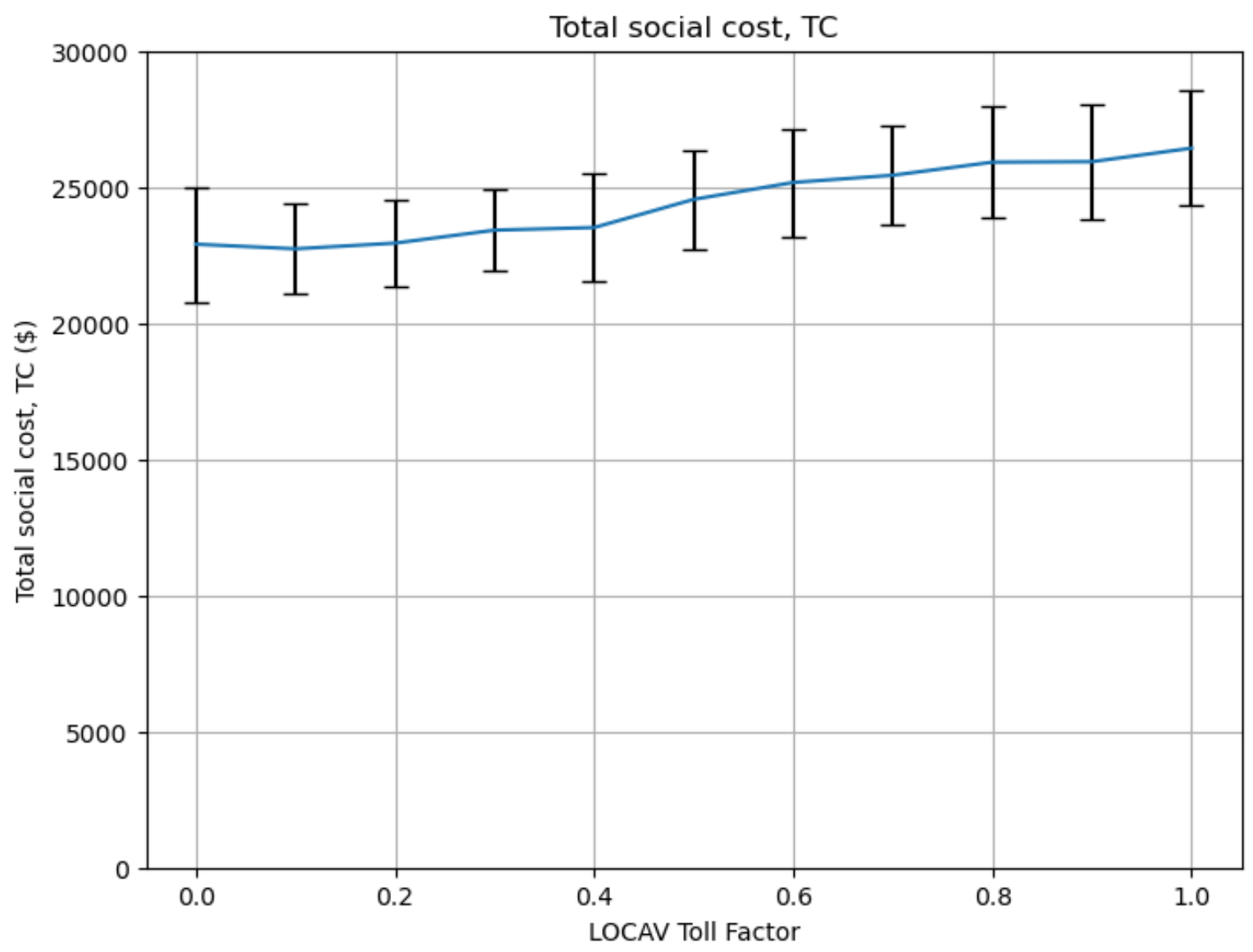

**Figure 14 - Total Social Cost against Low-Occupancy Connected and Automated Vehicle (LOCAV) Toll Factor**

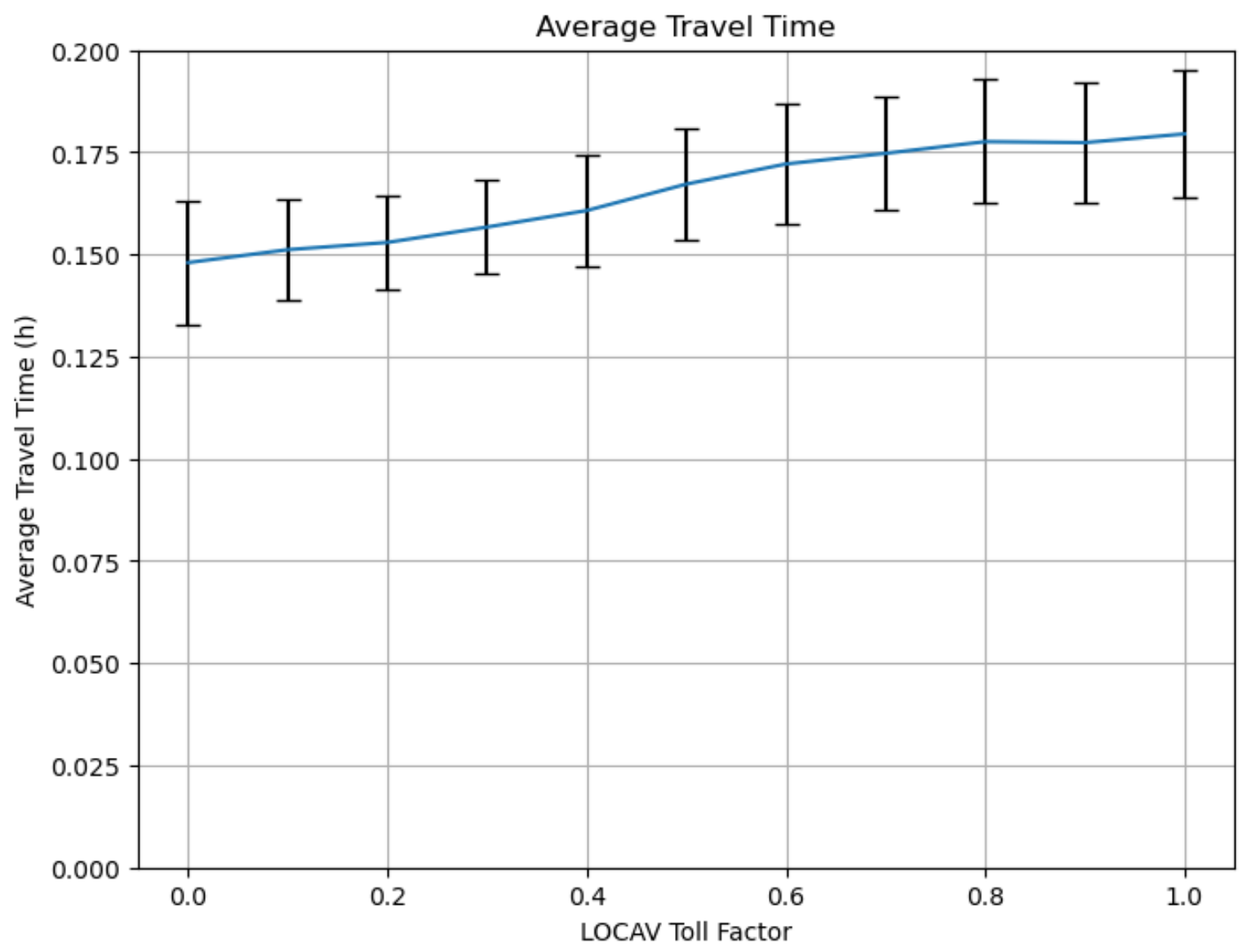

**Figure 15 - Average Travel Time against Low-Occupancy Connected and Automated Vehicle (LOCAV) Toll Factor**

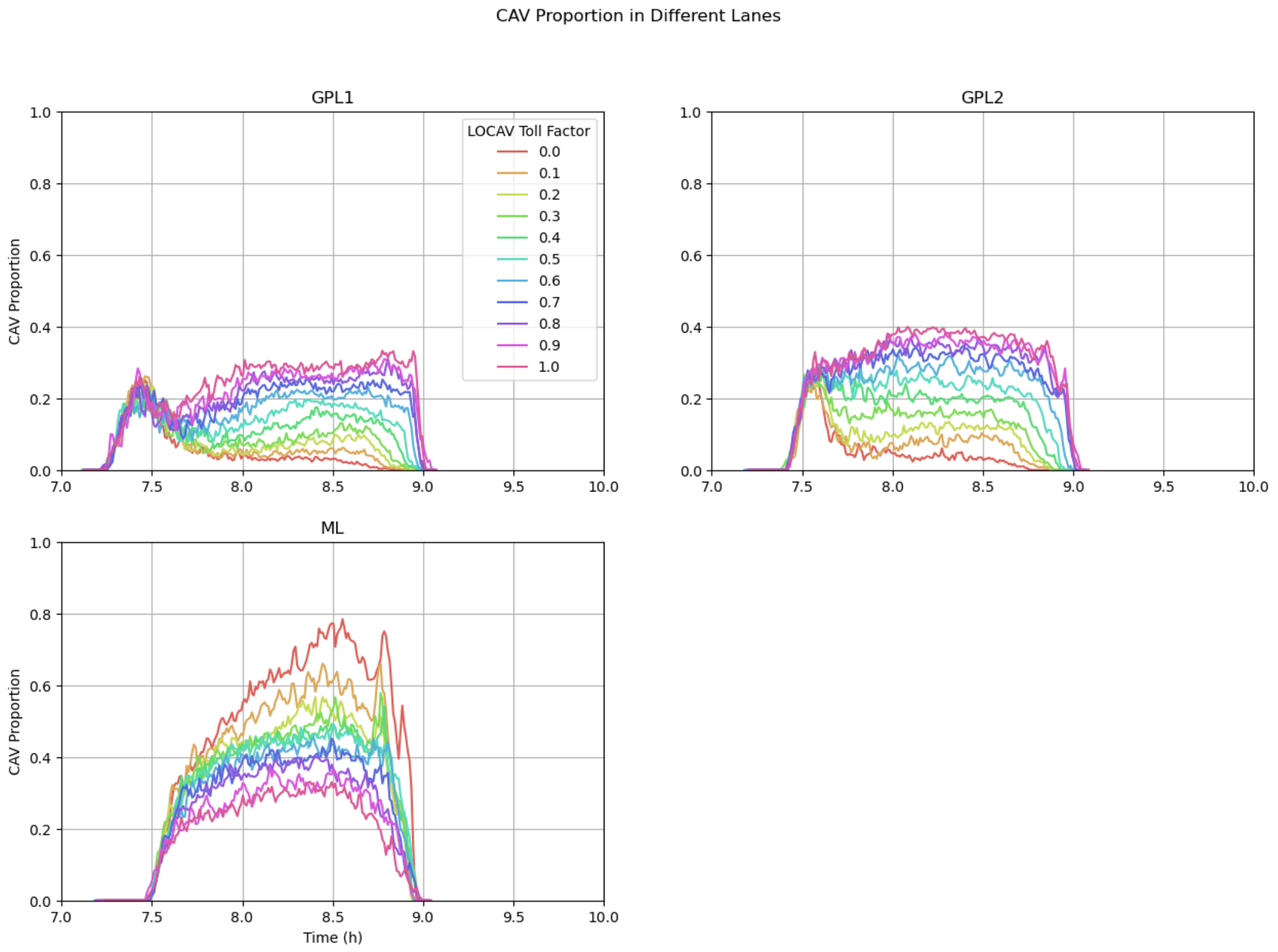

**Figure 16 - Connected and Automated Vehicle (CAV) Proportion in Different Lanes at Different Times**

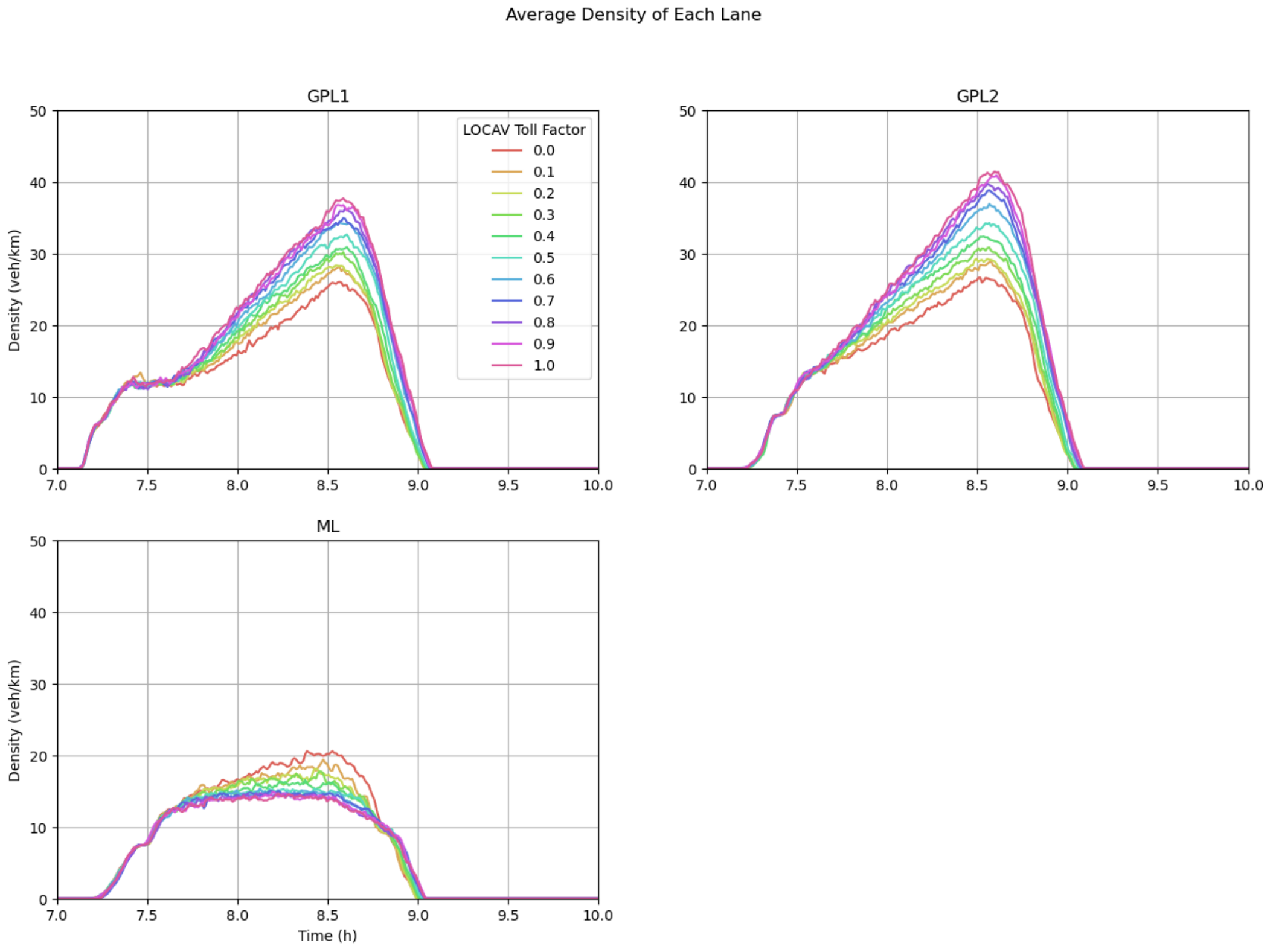

**Figure 17 - Average Traffic Density in Each Lane at Different Times**

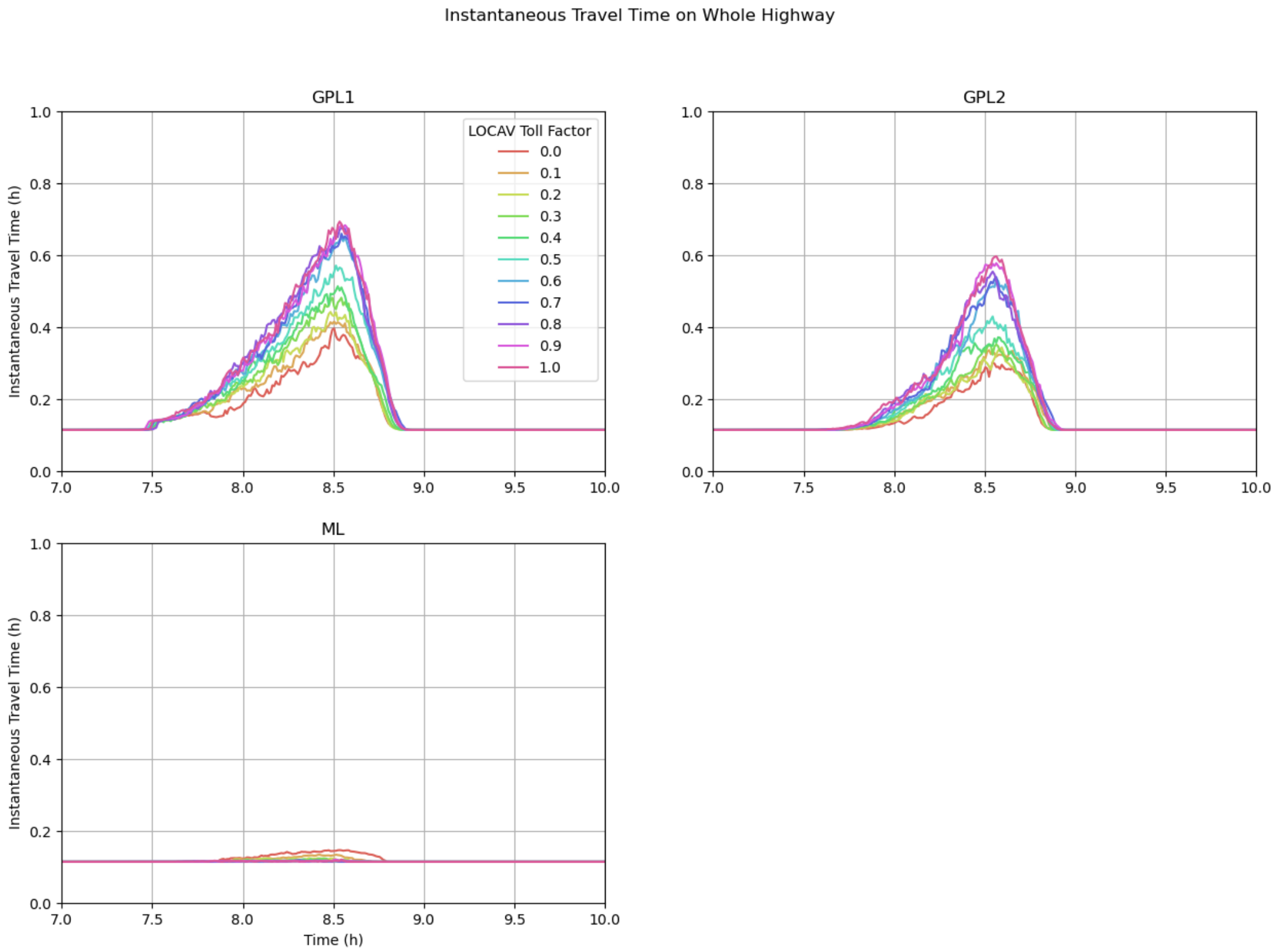

**Figure 18 - Instantaneous Travel Time on the Whole Highway at Different Times**

## 4.5 Sensitivity Analysis: High-Occupancy Connected and Automated Vehicle (HOCAV) Market Penetration Rate (MPR) - ML Policy ST1 (HOV/CAV use, HDV toll)

Following the discussion in Section 3.3, this sensitivity analysis evaluates the impacts of HOCAV MPRs on traffic in individual lanes. Figure 19 and Figure 20 show there is an average of $800 cost saving, or 0.33-minute time saving, per 10% conversion of HOHDVs to HOCAVs. This amounts to around $6.7 per HOCAV.

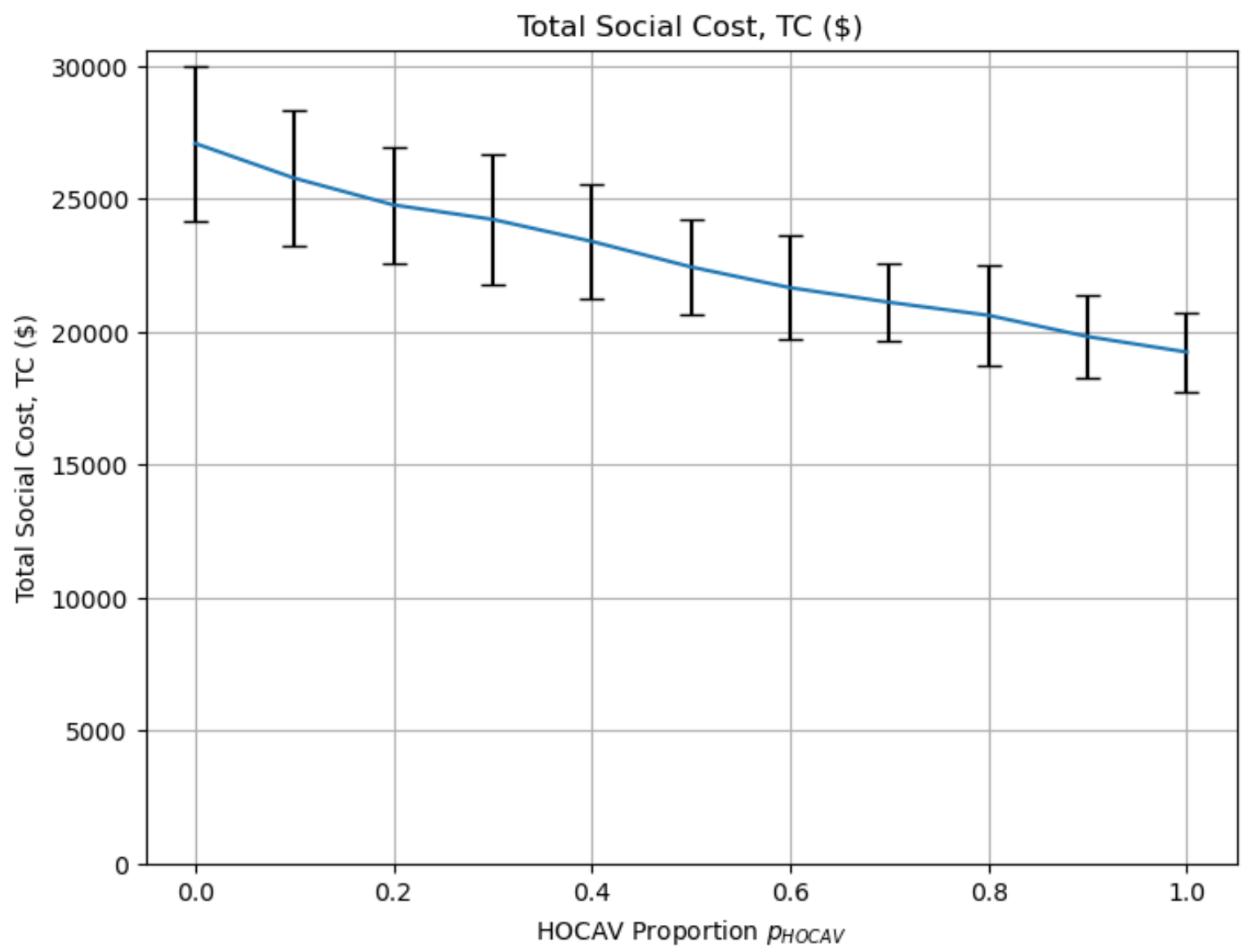

**Figure 19 - Total Social Cost against High-Occupancy Connected and Automated Vehicle (HOCAV) Market Penetration Rate (MPR)**

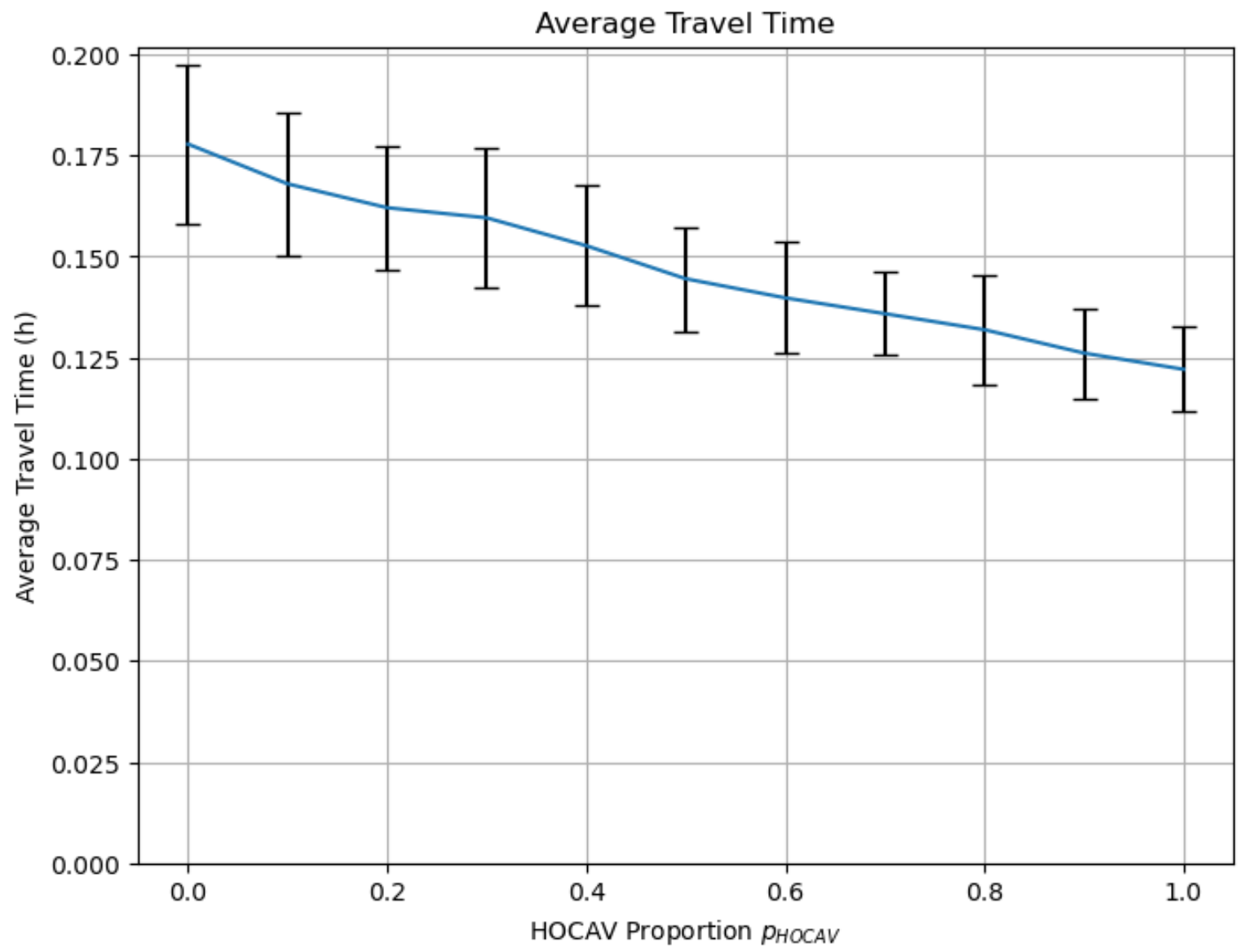

**Figure 20 - Average Travel Time against High-Occupancy Connected and Automated Vehicle (HOCAV) Market Penetration Rate (MPR)**

The savings are brought by vehicles shifted to the ML and remaining in GPLs. While not all additional CAVs stay in the ML (Figure 21) due to overcrowding or entries/exits, they considerably reduce the density and travel time in all three lanes (Figure 22 and Figure 23). Meanwhile, the CAV spillovers to GPLs also help improve their traffic.

The savings per HOCAV are smaller than the calculation results in Section 3.3 because not all HOCAVs converted travel in the ML along the highway at its busiest time. In other words, HOCAVs at free flows or in GPLs may not bring as much benefit as those in the ML at peak hours.

Moreover, as vehicles move faster in the ML, their proportion measured in density is diluted compared to slower vehicles.

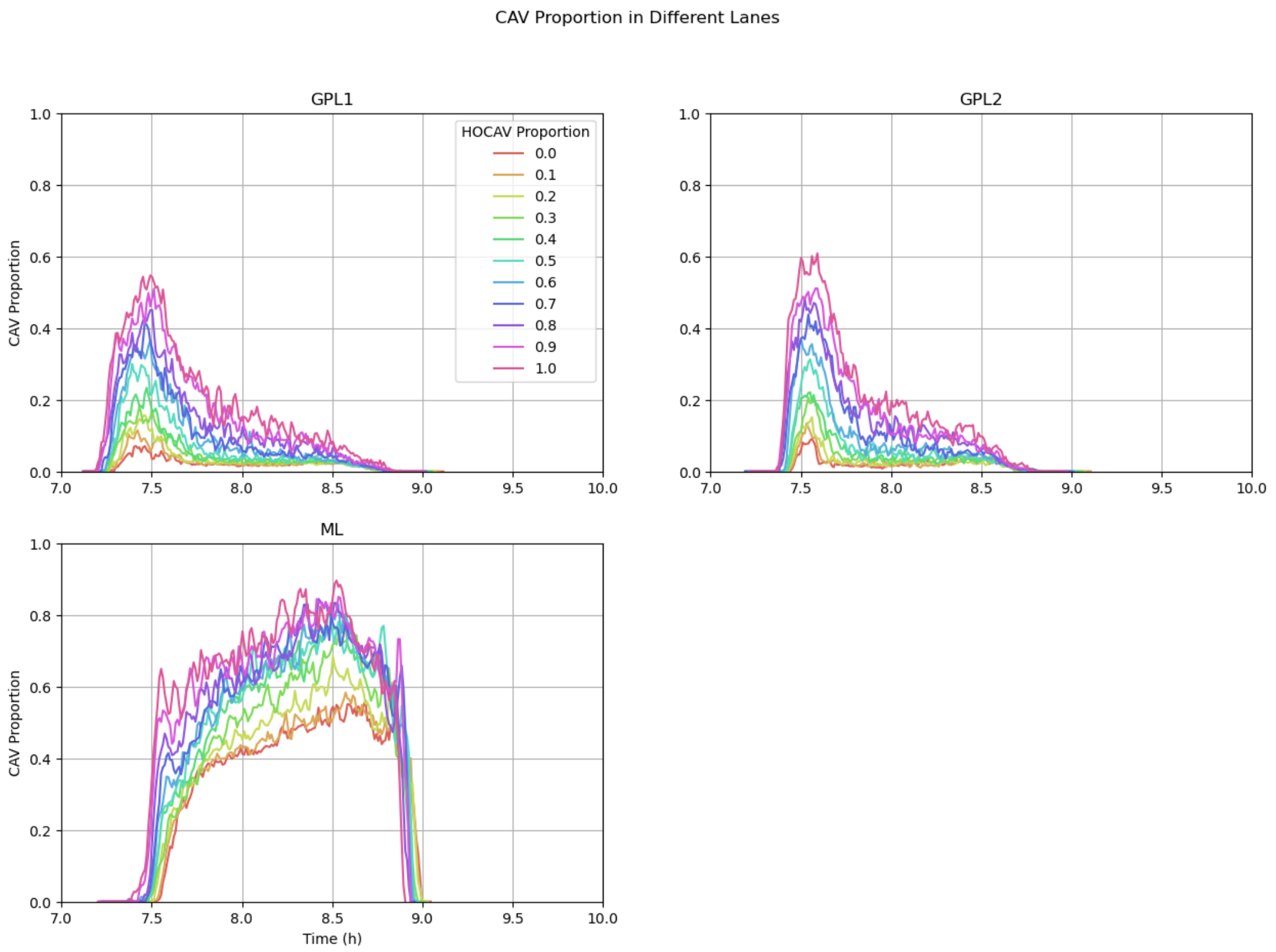

**Figure 21 - Connected and Automated Vehicle (CAV) Proportion in Different Lanes at Different Times**

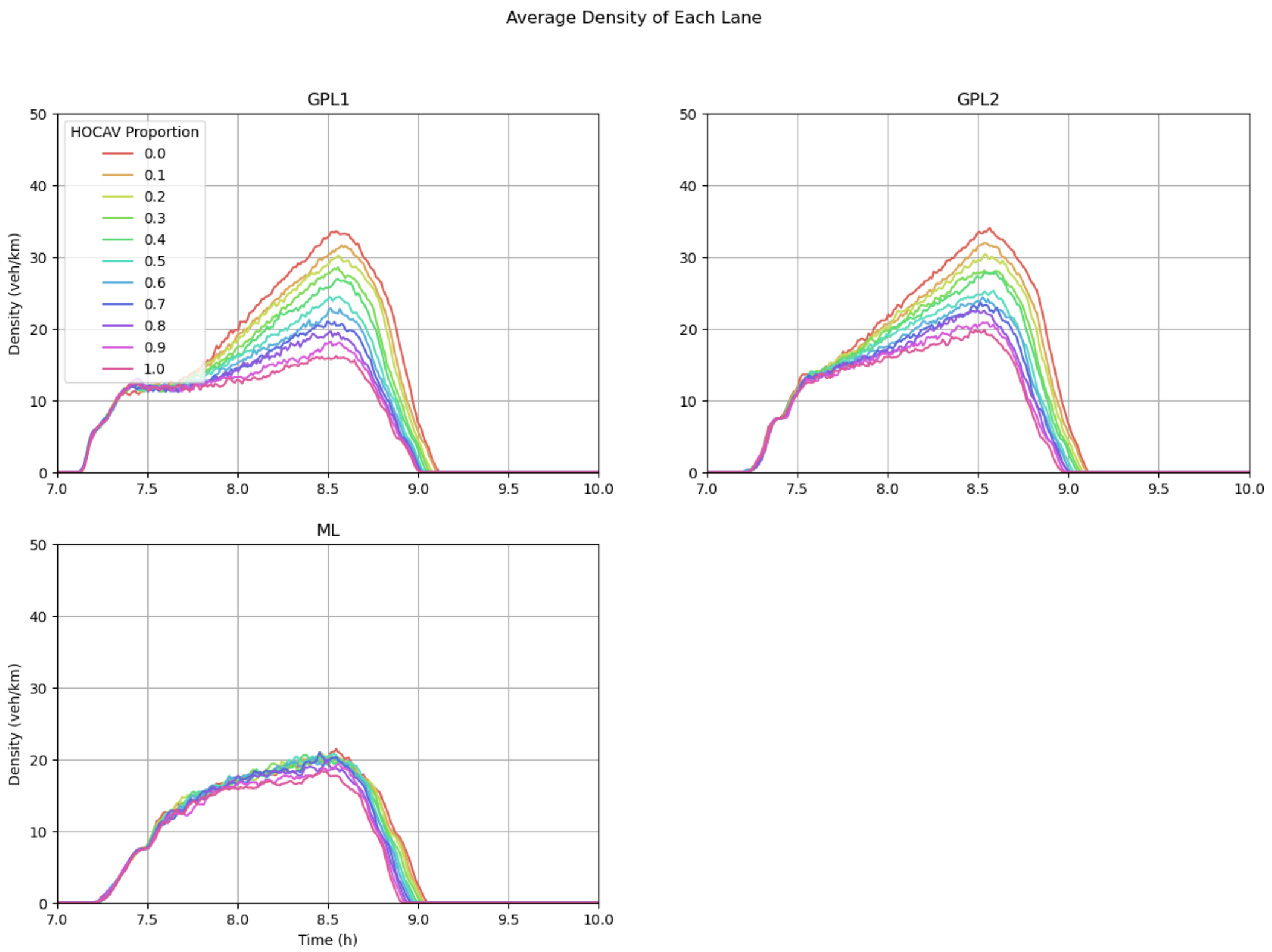

**Figure 22 - Average Traffic Density in Different Lanes at Different Times**

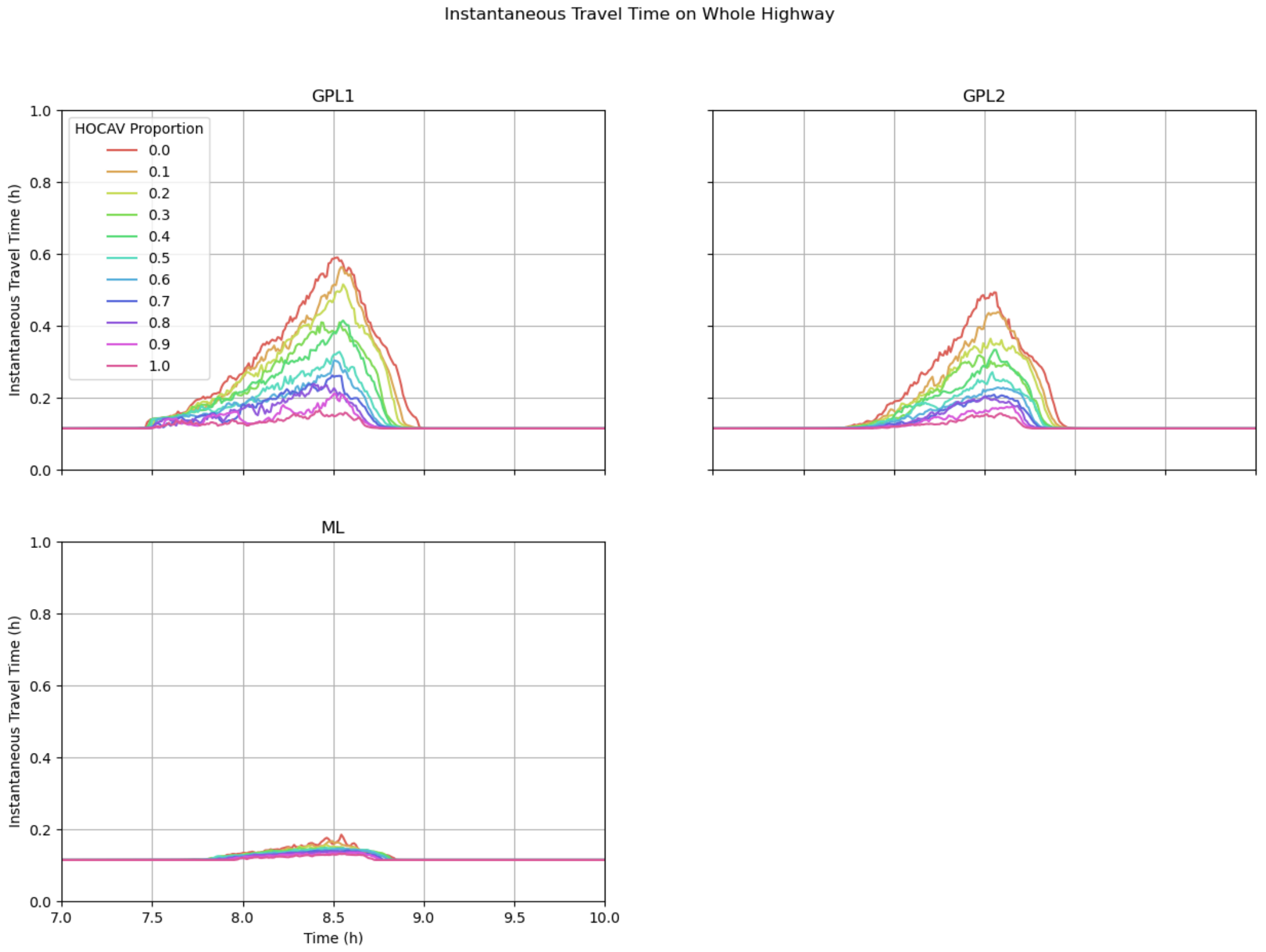

**Figure 23 - Instantaneous Travel Time on the Whole Highway in Different Lanes at Different Times**

# 5 CONCLUSION

This paper investigates different policies for ML usage and tolling on highways with mixed-traffic flows of HOVs, CAVs, and HDVs. A mesoscopic finite-difference model is developed to consider CAV/HDV flow-density relationship, lane-changing behavior, and multiple-entry/exit configuration. It shows that under limited CAV market penetration, exclusive lanes for HOVs and/or CAVs lead to excessive congestion in GPLs, which in turn blocks HOVs/CAVs from accessing the ML and leads to flow breakdown. A small and targeted toll for HDV ML use can avoid flow breakdown and allow higher-VOT vehicles to take advantage of the lower travel time, combining the traffic science of flow stability and the economics of resource allocation.

The simulation also illustrates the potential benefits of grouping CAVs and HOVs without a toll. As the simulation limits the MPRs of 2- and 3-passenger HOVs to 10% each, the actual benefits brought will be more significant in the presence of buses and other higher-VOT vehicles. Other benefits, including equity and emissions reductions, are also not considered.

Section 4.4 investigates the possibility of separating the CAV toll from HDVs. Results illustrate the difficulty in isolating the trade-off between extra capacity brought by CAVs and increased traffic. The optimal strategy to avoid flow breakdown and minimize social cost is possibly tolling HDVs before CAVs to maximize lane capacity. With the general increasing trend of cost and time over increasing toll, a natural follow-up question would be the possibility of a negative CAV toll to enhance concentration effects. Despite possible induced demand and additional difficulty, this may open up future avenues for investigation.

Sections 3.3 and 4.5 look into the potential of subsidizing HOHDV-HOCAV conversion for their positive externalities to the overall traffic. The benefits are evaluated and demonstrated based on different ratios of HDV:CAV for boosting the ML capacity to accommodate extra HDVs and alleviate GPL congestion. While the exact calculation depends heavily on the FD, as long as $\frac{\partial k_{i,l,t}^{cr}}{\partial n_{A,i,l,t}}$ is positive, there is a net benefit and positive externality, justifying incentives for HOHDV-HOCAV conversion for social good. However, further research is necessary for methods to target the critical HOHDVs which travel in the ML at peak hours and brings the most benefits upon conversion.

Seven of eight tolling policies assume no HOV toll, with the all-toll AT1 demonstrating no advantage over counterparts ST1 and ST2. An HOV toll violates the original intent to reduce traffic by encouraging ridesharing and public transit, which may further involve equity issues. However, a more interesting question is the effects on demand of improved traffic performance and change in the tolling regime. An endogenous demand model may be incorporated into this toll and traffic model for more insights.

The model leverages parameters which vary significantly according to highway characteristics, traffic patterns, and CAV/HDV interactions (in particular lane-changing algorithms) to highlight the underlying phenomenon. Precautions should be taken to analyze the results showcased in this work based on the presented road configuration and traffic parameters. Further sensitivity analysis should be conducted, or preferably tailor-made models under the proposed framework should be built for other highway settings and demand patterns, instead of directly extrapolating the results.

Nevertheless, this paper serves the proof-of-concept purpose of the novel combined usage/tolling regime under mixed traffic flows. It also develops a framework to assess the benefits of different tolling policies in different CAV/HOV MPRs, with the flexibility to incorporate more precise mixed-traffic FD (e.g., non-linear) and tolling mechanism (e.g., anticipatory). Future research can look into applying the model to data of existing highways with ML/HOV/HOT systems and conduct more extensive sensitivity analysis on parameters including mixed traffic flow behavior.

## AUTHOR CONTRIBUTIONS

All authors contributed to the study design, analysis and interpretation of results, and manuscript preparation. All authors reviewed the results and approved the submission of the manuscript.